\documentclass[11pt,a4paper]{article}
\pdfoutput=1
\usepackage{amsmath}
\usepackage{amsfonts}
\usepackage[normalem]{ulem}
\usepackage{soul}
\usepackage{color}
\usepackage[thicklines]{cancel}
\usepackage{graphicx}
\usepackage{dcolumn}
\usepackage{bm}
\usepackage[
bookmarks,bookmarksnumbered,colorlinks=true,anchorcolor=blue,
linkcolor=blue,urlcolor=blue,citecolor=blue,
breaklinks=true,hypertexnames=false]{hyperref}
\usepackage{geometry}
\usepackage{authblk}
\usepackage{amssymb}
\usepackage[utf8]{inputenc}
\usepackage[numbers,sort&compress]{natbib}
\usepackage{pifont}
\usepackage{tikz}
\usepackage{eso-pic}
\usepackage{multirow}
\usepackage{float}
\numberwithin{equation}{section}   
\usepackage{booktabs}
\usepackage{lmodern}
\usepackage{microtype}

\def\cO{\mathcal{O}}
\def\cP{\mathcal{P}}

\def\cV{\mathcal{V}}

\newcommand{\bea}{\begin{eqnarray}}   \newcommand{\eea}{\end{eqnarray}}

\begin{document}

\AddToShipoutPictureFG*{%
  \AtPageUpperLeft{%
    \put(0,0){\makebox[\paperwidth][r]{%
      \hspace*{-2.6cm}%
      \raisebox{-1.6cm}[0pt][0pt]{\small\texttt{IFT-UAM/CSIC-26-109}}%
      \hspace*{2.6cm}%
    }}%
  }%
}

\title{
 Fibre Inflation Beyond BHP: \\
 Constraints and a BHP-Free Regime}
 
 \author[1]{Xin Gao
 	\footnote{xingao@scu.edu.cn}}
 \author[1]{Zhiwei Li
 	\footnote{lizhiwei4@stu.scu.edu.cn}}
 \author[2]{Chuying Wang
 	\footnote{chuying.wang@ift.csic.es}}
 
 \affil[1]{College of Physics, Sichuan University, Chengdu 610065, China}
 \affil[2]{Instituto de F\'isica Te\'orica UAM/CSIC, Universidad Aut\'onoma de Madrid, 28049 Madrid, Spain}
 
 \date{}

 \maketitle

\begin{abstract}

We study a restricted fibre-dependent \(\alpha'\)-loop sector in Type IIB
Large Volume fibre inflation beyond the conventional
Berg--Haack--Pajer (BHP) parametrisation. Motivated by
effective-field-theory corrections in K3-fibred Calabi--Yau
orientifolds, we analyse BHP-complete, partially BHP-driven and
BHP-free configurations. On the BHP-connected branches, the additional loop coefficients are bounded
at the percent level relative to their matched BHP coefficients, with the
logarithmic coefficient typically more tightly constrained by its effects
on the slope and curvature. 
At the effective-field-theory level, we further identify a BHP-free
fibre-inflation regime, which we refer to as \(\alpha'\)-loop inflation,
in which the conventional KK- and winding-type BHP channels vanish while
the generalized loop sector stabilises the fibre modulus and supports
slow-roll inflation. This regime exhibits a characteristic hierarchy
between the different fibre-dependent loop structures, controlled by
the large-volume suppression, and satisfies CMB constraints,
K\"ahler-cone control, large-volume expansion and heavy-modulus stability. Our results show that
fibre inflation is sensitive to the orientifold and brane configurations, and that corrections
beyond the BHP ansatz can generate new controlled inflationary regimes.

\end{abstract}

\clearpage

\tableofcontents

\section{Introduction}

String compactifications provide a framework in which inflationary dynamics can be related to the geometry and quantum corrections of extra dimensions. In Type IIB Calabi--Yau orientifolds, background three-form fluxes generate the Gukov--Vafa--Witten superpotential and can stabilise the axio-dilaton and complex-structure moduli, while the K\"ahler moduli remain unfixed at tree level because of the no-scale structure \cite{Gukov:1999ya,Giddings:2001yu}. Their stabilisation therefore relies on perturbative and non-perturbative effects. Two prominent mechanisms are the KKLT construction \cite{Kachru:2003aw} and the Large Volume Scenario (LVS) \cite{Becker:2002nn,Balasubramanian:2005zx,Conlon:2005ki}. In the LVS, the leading \(\alpha'^3\) correction to the K\"ahler potential is balanced against non-perturbative contributions to the superpotential, stabilising the compactification at exponentially large volume and providing a controlled setting for string cosmology.

K\"ahler moduli that remain light after leading stabilisation are natural inflaton candidates. This idea underlies several classes of LVS inflation, such as blow-up inflation \cite{Conlon:2005jm}, which is driven by a blow-up mode lifted by non-perturbative corrections; poly-instanton inflation \cite{Cicoli:2011ct,Blumenhagen:2012kz,Blumenhagen:2012ue,Lust:2013kt} instead is generated by poly-instanton effects, with the inflaton associated with a rigid four-cycle carrying exact Wilson lines. The focus of this paper is fibre inflation \cite{Cicoli:2008gp,Krippendorf:2009zza,Cicoli:2013oba,Burgess:2016owb,Cicoli:2016xae,Cabella:2017zsa,Kallosh:2017wku,Cicoli:2017axo}, which is realised in K3- or \(T^4\)-fibred orientifold Calabi--Yau compactifications. The leading LVS potential fixes the overall volume \(\mathcal V\) and a blow-up modulus, while leaving a comparatively light fibre modulus \(\tau_f\).
Subleading string-loop and higher-derivative corrections then lift this flat direction and generate the inflationary potential. 
However, such a construction makes the scenario particularly sensitive to various corrections to the K\"ahler potential and, crucially, to the D-brane/orientifold configuration. A correction
that is harmless for the stabilisation of the heavy moduli may therefore compete directly
with the fibre potential.

The standard description of loop effects in fibre inflation is based on the one-loop K\"ahler-potential corrections computed in toroidal orientifolds \cite{Berg:2005ja} and conjecturally extended to Calabi--Yau orientifolds in \cite{Berg:2007wt}, known as the BHP conjecture. In this parametrisation, Kaluza--Klein-type and winding-type corrections depend on the details of D-brane/O-plane configurations and on the volumes of the cycles exchanged by closed strings \cite{Berg:2007wt,Cicoli:2007xp}.
KK-type terms require suitable transverse KK exchange channels, including non-intersecting D7/O7 sources or O3--D7/O7 exchange, whereas winding-type terms require intersecting D7/O7 sources whose intersection curve has non-trivial one-cycles. Thus the BHP contribution can be present, partially absent, or vanish altogether depending on the orientifold and D-brane setup.  Recent progress in the systematic classification of toric
Calabi--Yau orientifolds and the construction of large-scale
orientifold databases
\cite{Gao:2013pra,Altman:2021pyc,Crino:2022zjk,Cao:2024oqx}
provides increasingly detailed geometric input for analysing the
O-plane content and D-brane configurations relevant to these different
BHP channels.

The BHP ansatz has been central to the development of fibre inflation, explicit global embeddings
\cite{Cicoli:2016xae,Cicoli:2017axo,Bera:2024emb,Cicoli:2024bxw},
and other extensions 
\cite{Leontaris:2025afi,Chakraborty:2025wqn}, but it need not exhaust the perturbative corrections of a generic Calabi--Yau orientifold. Recently, a higher-dimensional effective-field-theory analysis distinguishes three physically different sources: classical warping, finite non-local genuine loop effects analogous to Casimir energies, and local higher-derivative operators in the bulk or on the D-branes/O-plane system \cite{Gao:2022uop}. The local terms may be present as Wilsonian operators or as counterterms required by the ultraviolet region of loop diagrams. In particular, marginal local operators on D-brane systems can generate logarithmic dependence on compactification radii. Recent work on loop blow-up inflation has also shown that genuine loop effects can reorganise an inflationary potential rather than merely spoil it \cite{Bansal:2024uzr}. Meanwhile, higher-derivative effects descending from the ten-dimensional \(\alpha'^3R^4\) sector give additional $F^4$ scalar-potential contributions \cite{Ciupke:2015msa,Cicoli:2016chb,Cicoli:2023njy}.
  These developments motivate a reanalysis of fibre inflation without assuming that the BHP parametrisation is complete.

In this work, the expression ``$\alpha'$-loop corrections'' is used in a
restricted fibred-geometry sense. It denotes the fibre-dependent inverse-radius sector suggested by the two-step EFT analysis of K3-fibred compactifications, rather than the full set of
perturbative corrections in Type IIB orientifolds. After the axio-dilaton and complex-structure moduli are stabilised, we denote by $W_0$ the flux superpotential, by $\mathcal V$ the Einstein-frame Calabi--Yau volume, and use the convention $\kappa=g_s/(8\pi)$. The resulting ``$\alpha'$-loop corrections''  to the scalar potential are parametrised as
\begin{equation}
V_{\alpha'\text{-loop}}
=
\kappa\frac{|W_0|^2}{\mathcal V^2}
\left[
\frac{\theta_1}{\tau_f^2}
-
\frac{
\theta_2+
\theta_3\left(\ln \tau_f-\frac32\right)
}
{\mathcal V\sqrt{\tau_f}}
\right].
\label{eq:intro-alpha-loop-potential}
\end{equation}
where \(\theta_1\) parametrises the fibre-dependent degree-\((-2)\)
structure in the four-cycle volumes, including genuine non-local fibre-loop
effects and possible  brane-localized genuine loops.
 \(\theta_2\) parameterises the 
 mixed D7/O7-sector contribution, including threshold pieces and also possible
brane-localized genuine loops  with the same moduli dependence. 
\(\theta_3\) controls the logarithmic enhancement 
associated with candidate marginal  operators on brane systems. 
In principle, the coefficients should be computable from the microscopic string compactification data, including the Calabi-Yau metric and relevant string amplitudes. Since such microscopic data are not explicitly available for the
compactifications studied here,
we treat them as effective parameters of moduli structures and constrain them by perturbative control, heavy-modulus stability and inflationary flatness.

At the level of the four-dimensional scalar potential, 
the \(\theta_1\) term multiplies
the decreasing \(1/\tau_f^2\) structure that can occur in the KK-like BHP
correction, whereas \(\theta_2\) multiplies the winding-like BHP mixed
structure \(1/(\mathcal V\sqrt{\tau_f})\). The \(\theta_3\) term gives
a logarithmically running deformation of the same mixed structure.
When the BHP terms are present, only the effective combinations are visible in the scalar potential; our bounds should therefore be understood as constraints on additional effective deformations of a BHP baseline. When the conventional BHP coefficients vanish because the required transverse or intersection channels are absent, the corrections in \eqref{eq:intro-alpha-loop-potential} can nevertheless survive and may become a leading ingredient of the fibre potential.  Their coefficients nevertheless remain sensitive to the orientifold projection, the local D7/O7 geometry and the wrapped divisors. This distinction leads to three BHP baselines studied for
BHP-driven inflation: BHP-complete, KK-only and winding-only.
If both KK and winding channels vanish, one obtains the separate
BHP-free regime.

The central questions are then the following. How large can the additional fibred-geometry corrections be before they destroy the fibre-inflation plateau? How does the answer depend on whether the brane configuration generates KK-type, winding-type, or no conventional BHP terms? Can the corrections in \eqref{eq:intro-alpha-loop-potential} themselves stabilise the fibre direction and support slow-roll inflation in a BHP-free configuration? If so, can the resulting single-field description be made compatible with the stability of the heavy LVS volume and blow-up moduli? 
Addressing these questions requires both the effective potential and
explicit orientifold Calabi--Yau data, since the allowed correction
pattern is determined by the orientifold involution, O-plane content
and D-brane configuration.

Our main results are as follows:
\begin{itemize}
\item
 We identify a novel BHP-free fibre-inflation regime driven by the generalized
$\alpha'$-loop sector. In an explicit orientifold setup with absent conventional KK- and
winding-type BHP channels, we show at the effective  field theory level that the generalized corrections \eqref{eq:intro-alpha-loop-potential}  can nevertheless
stabilise the fibre modulus and support slow-roll inflation, both with and without the higher-derivative $F^4$ contribution.
In the leading large-volume  approximation, the fibre minimum satisfies
\begin{equation}
    \frac{\theta_1}{\theta_2}
=
\frac{\langle\tau_f\rangle^{3/2}}{4\cV}\left(1+ \frac{\theta_3}{\theta_2}(\ln \langle\tau_f \rangle- \frac{7}{2})\right).
\label{eq:theta1theta2}
\end{equation}
This relation shows that the BHP-free regime is not obtained by an
arbitrary tuning of independent coefficients, but is controlled by the
large-volume hierarchy of the fibre geometry.
This gives a qualitatively new fibre-inflation mechanism, which we
refer to as \(\alpha'\)-loop fibre inflation, beyond the standard BHP construction. 

\item The viability of the BHP-free fibre-inflation regime is controlled by several constraints at once. The valid inflation scenario, the CMB amplitude, the K\"ahler-cone range, the LVS stabilisation condition, and the hierarchy between the Hubble scale and the heavy moduli combine to give a finite allowed window for  $(\theta_1,\theta_2, \theta_3)$. 
The
logarithmic coefficient $\theta_3$ is treated as an independent shape parameter rather than assumed
to be subleading.  Since it affects the potential, slope and curvature, in the favourable regions of this window, 
controlled BHP-free inflation can be realised with 
\begin{equation}
 \theta_2 \simeq \cO(10^{-2}), \qquad |\frac{\theta_3}{\theta_2}|\lesssim \cO(10^{-2})\text{--}\cO(10^{-1}),
\end{equation}
with the precise bound depending on the trajectory and on the desired level of hierarchy. In viable solutions within this window, the logarithmically enhanced
contribution can be comparable to or larger than the non-logarithmic
$\theta_2$ term at horizon exit. Together with \eqref{eq:theta1theta2}, $\theta_1$ is bounded by the characteristic hierarchy for generic values of $\langle\tau_f\rangle \sim \cO(1\text{--}100)$:
\begin{equation}
     \frac{\theta_1}{\theta_2}
\simeq \cO(1)
\frac{\langle\tau_f\rangle^{3/2}}{4\cV}
\end{equation}

\item
When conventional BHP terms generate the leading fibre potential, the \(\alpha'\)-loop sector is bounded by the requirement that the standard fibre-inflation plateau remain viable. We analyse both BHP-complete
setups, where KK- and winding-type contributions are present, and partially BHP-driven setups, where one of these channels is absent. In these cases, the non-logarithmic \(\theta_1\) and \(\theta_2\)
contributions project onto the same single-field power-law structures as the corresponding BHP terms, while \(\theta_3\) gives a genuine logarithmic deformation. On the standard BHP-connected branch, the upper bounds on
the corresponding fractional deformations are \(\mathcal O(10^{-2})\) compared to the original BHP coefficients, reaching at most
a few percent in the least restrictive cases.

If only one BHP channel is present, the correction associated with the
missing structure need not be small. In the KK-only case,
\(\theta_2\) and \(\theta_3\) can supply the missing winding-like
mixed term, whereas in the winding-only case \(\theta_1\) can provide
the missing \(1/\tau_f^2\) structure. The allowed raw coefficients are
therefore model dependent; the physically relevant constraints are
their effects on the full potential, its first two derivatives and the
stability of the heavy moduli.

\item
Our analysis shows that the fibre potential is not fixed by the Calabi--Yau volume form alone. The orientifold involution and D-brane/O-plane arrangement determine which BHP channels are present, and hence whether the generalised $\alpha'$-loop corrections act as small deformations, complement a partially reduced BHP potential, or become the leading source of the inflaton potential.
\end{itemize}

This paper is organised as follows. Section~\ref{sec:bg} reviews LVS moduli stabilisation, fibre inflation and the perturbative corrections relevant to the four-dimensional effective action. Section~\ref{sec:BHP-deforme} derives the fibred-geometry \(\alpha'\)-loop potential and quantifies in explicit K3-fibred geometries how such broader correction patterns constrain BHP-generated fibre inflation. Section~\ref{sec:alpha-loop-inflation} analyses the BHP-free regime in which inflation is driven by the \(\alpha'\)-loop sector, including the joint constraints from the CMB normalisation, K\"ahler-cone field range and heavy-modulus hierarchy. Section~\ref{sec:con} summarises our results and discusses further directions. To avoid confusion caused by the different frame choices and superpotential normalisations used in the literature, Appendix~\ref{frame} fixes our conventions and explains how they are related. Appendix~\ref{app:alpha_loop_origin} reviews the origin of the terms appearing in the fibred \(\alpha'\)-loop correction.

\section{LVS Fibre Inflation and Perturbative Corrections}
\label{sec:bg}

In this section we first review the LVS stabilisation of the heavy K\"ahler moduli and the standard fibre-inflation potential and then summarise the perturbative corrections relevant for our analysis. Throughout this paper, we work in the standard four-dimensional
Einstein frame, set $M_p=1$, denote the dimensionless Einstein-frame
Calabi--Yau volume by $\mathcal V$, and use
$\kappa=g_s/(8\pi)$. Our superpotential and frame conventions are summarised in
Appendix~\ref{frame}.

\subsection{LVS stabilisation and fibre inflation}
\label{LVS}

In Type IIB Calabi--Yau orientifold compactifications, background three-form fluxes generate the Gukov--Vafa--Witten superpotential and stabilise the axio-dilaton and complex-structure moduli, while the K\"ahler moduli remain unfixed at tree level because of the no-scale structure \cite{Gukov:1999ya,Giddings:2001yu}. After the heavy complex-structure and dilaton fields are integrated out, the K\"ahler sector is described by the $\mathcal N=1$ F-term potential:
\begin{equation}
    V_F=\kappa e^K\left(K^{I\bar J}D_IW D_{\bar J}\overline W-3|W|^2\right),
    \qquad
    D_IW=\partial_IW+W\partial_IK .
    \label{eq:F-term-potential}
\end{equation}

In the Large Volume Scenario (LVS), the leading perturbative $\alpha'^3$ correction to the K\"ahler potential is balanced against non-perturbative superpotential terms \cite{Becker:2002nn,Balasubramanian:2005zx,Conlon:2005ki}. Schematically,
\begin{equation}
    K=-2\ln\left(\mathcal V+\frac{\hat\xi}{2}\right)+\cdots,
    \qquad
    W=W_0+\sum_i A_i e^{-a_iT_i},
    \qquad
    \hat\xi=\frac{\xi}{g_s^{3/2}},
    \label{eq:LVS-K-W}
\end{equation}
where $\xi = - \frac{\chi({X}_3)\zeta(3)}{2(2\pi)^3}$,  $\zeta(3) \simeq 1.2$ is the Riemann zeta function and $\chi({X}_3)$ is the Euler characteristic of the Calabi--Yau threefold ${X}_3$.

Fibre inflation is realised in a weak Swiss-cheese geometry containing a K3- or \(T^4\)-fibred structure. A convenient parametrisation is \cite{Cicoli:2009zh}
\begin{equation}
    \mathcal V
    =
    \alpha\left(\sqrt{\tau_f}\,\tau_b-\gamma\tau_s^{3/2}\right),
    \qquad
    \tau_b\gg \tau_f >\tau_s,
    \label{eq:fibred-volume-review}
\end{equation}
where  $\alpha, \gamma$ are positive model-dependent constants and \(\tau_f\) is the fibre modulus, \(\tau_b\) controls the base direction and \(\tau_s\) is the blow-up modulus. The leading LVS scalar potential takes the form
\begin{equation}
V_{\rm LVS} (\mathcal{V} ,\, \tau_s)
=
\frac{
8\kappa\sqrt{\tau_s}\,(a_s A_s)^2 e^{-2a_s\tau_s}
}{
3\alpha\gamma\mathcal V
}
-
\frac{
4\kappa W_0 a_s A_s\tau_s e^{-a_s\tau_s}
}{
\mathcal V^2
}
+
\frac{
3\kappa\hat{\xi}W_0^2
}{
4\mathcal V^3
}.
\label{eq:LVS-potential}
\end{equation}
The stationarity condition with respect to the blow-up modulus,
\(\partial_{\tau_s}V_{\rm LVS}=0\), can be solved algebraically
for the overall volume, giving
\begin{align}
\left\langle\mathcal V\right\rangle
&=
\frac{
3\alpha\gamma
\sqrt{\left\langle\tau_s\right\rangle}
\,|W_0|
\left(1-a_s\left\langle\tau_s\right\rangle\right)
}{
a_sA_s
\left(1-4a_s\left\langle\tau_s\right\rangle\right)
}
e^{a_s\left\langle\tau_s\right\rangle},
\label{eq:LVS-minimum-volume}
\end{align}
where \(a_s=2\pi\) for a Euclidean D3-brane instanton and
\(a_s=2\pi/N\) for gaugino condensation in an \(SU(N)\) gauge theory.
The stationarity condition with respect to the overall volume,
\(\partial_{\mathcal V}V_{\rm LVS}=0\), together with
\eqref{eq:LVS-minimum-volume} yields the approximate minimum in the
large \(a_s\tau_s\) regime:
\begin{equation}
\left\langle\tau_s\right\rangle
\simeq
\left(
\frac{\hat{\xi}}{2\alpha\gamma}
\right)^{2/3},
\qquad \left\langle\mathcal V\right\rangle
\simeq
\frac{
3\alpha\gamma
\sqrt{\left\langle\tau_s\right\rangle}
\,|W_0|
}{
4a_sA_s
}
e^{a_s\left\langle\tau_s\right\rangle},
\qquad
a_s\left\langle\tau_s\right\rangle\gg 1 .
\label{eq:LVS-tau-asymptotic}
\end{equation}
Although these approximate expressions provide useful analytic intuition, the relations in \eqref{eq:LVS-tau-asymptotic} are reliable only when
\(a_s\left\langle\tau_s\right\rangle \gg 1\). For the sake of conservatism, all
numerical results below are obtained from the unexpanded LVS
stationarity conditions, rather than the above asymptotic expressions.

The leading LVS potential fixes \(\mathcal V\) and \(\tau_s\) but leaves the fibre modulus comparatively light. The fibre potential is therefore generated only by subleading perturbative effects, making fibre inflation especially sensitive to loop and higher-derivative corrections.
We now recall how a small correction \(\delta K\) to the no-scale K\"ahler potential contributes to the scalar potential. Let
\begin{equation}
    K=K_0+\epsilon\,\delta K,
    \qquad
    K_0=-2\ln\mathcal V,
    \qquad
    W=W_0,
    \label{eq:K-expansion}
\end{equation}
and expand the scalar potential as
$  V_F
    =
    \delta V_0+\epsilon\,\delta V_1+
    \epsilon^2\delta V_2+\mathcal O(\epsilon^3).
$
The no-scale structure gives \(\delta V_0=0\), and the linear correction is
\begin{equation}
    \delta V_1
    =
    -\kappa
    \left(
       2\tau_i\frac{\partial\delta K}{\partial\tau_i}
       +
       \tau_i\tau_j
       \frac{\partial^2\delta K}{\partial\tau_i\partial\tau_j}
    \right)
    \frac{|W_0|^2}{\mathcal V^2} .
    \label{eq:linear-deltaV}
\end{equation}
If \(\delta K\) is homogeneous of degree \(n\) in four-cycle volumes, then
\begin{equation}
    \delta V_1
    =
    -\kappa\, n(n+1)\,
    \delta K\,
    \frac{|W_0|^2}{\mathcal V^2} .
    \label{eq:homogeneity-deltaV}
\end{equation}
This observation is important in what follows. For degree \((-1)\) corrections, \(n(n+1)=0\), so the linear term vanishes by the extended no-scale structure and the first non-zero contribution arises at quadratic order in \(\delta K\). This is the standard behaviour of KK-type BHP corrections and of local relevant D7/O7 operators in the four cycle volume convention used here. By contrast, degree \((-2)\) corrections contribute already through \(\delta V_1\). The \(\alpha'\)-loop fibre potential derived in Section~\ref{sec:alpha-loop-correction} belongs to this second class.

The conventional description of string-loop effects is the BHP ansatz, motivated by explicit toroidal-orientifold calculations and conjectured to extend to Calabi-Yau orientifolds \cite{Berg:2005ja,Berg:2007wt,Cicoli:2007xp,Cicoli:2008va,Cicoli:2009zh}. It consists of two types of corrections to the K\"ahler potential,
\begin{equation}
    \delta K^{\rm KK}_{g_s}
    \sim
    \sum_a
    \frac{g_s\, C^{\rm KK}_a\,{\cal T}_a(t^\perp)}
         {\mathcal V},
    \qquad
    \delta K^{\rm W}_{g_s}
    \sim
    \sum_a
    \frac{C^{\rm W}_a}
         {{\cal I}_a(t^\cap)\,\mathcal V},
    \label{eq:BHP-K}
\end{equation}
where ${\cal T}_a(t^\perp)$ and ${\cal I}_a(t^\cap)$ are, in the toroidal
case, linear in two-cycle volumes. The superscript $\perp$
indicates a transverse two-cycle, while $\cap$ denotes a two-cycle
associated with an intersection locus. It was shown in {\cite{Gao:2022uop}} that the same homogeneous scaling may
survive in generic Calabi--Yau geometries, although the functions need not be strictly linear in the two-cycle volumes.

According to (\ref{eq:homogeneity-deltaV}), the KK-type correction is homogeneous of degree
$-1$ in four-cycle variables and its linear contribution to the scalar
potential cancels. Its leading effect is therefore quadratic in
$\delta K^{\rm KK}_{g_s}$. By contrast, the winding-type correction is
homogeneous of degree $(-2)$ and contributes already at linear order.
For a simple swiss-cheese K3-fibred volume
\(
\mathcal V\simeq \frac12\sqrt{\tau_f}\tau_b
\),
the BHP-induced fibre potential takes the standard form
\begin{equation}
    V_{\rm BHP}
    =
    \kappa\frac{|W_0|^2}{\mathcal V^2}
    \left(
        \frac{A}{\tau_f^2}
        -
        \frac{B}{\mathcal V\sqrt{\tau_f}}
        +
        \frac{C\,\tau_f}{\mathcal V^2}
    \right),
    \label{eq:BHP-potential}
\end{equation}
where the $A, C$ terms arise from KK-type corrections 
proportional to $(g_s C^{\rm KK})^2$, while the $B$ term comes
from winding-type corrections proportional to $C^{\rm W}$. 
Hence the KK-type corrections are suppressed by a factor 
$g_s^2$. 
The coefficients
$A$, $B$ and $C$ encode the detailed dependence on the D-brane/O-plane configuration in a concrete Calabi--Yau geometry and are treated as effective parameters constrained by consistency with the inflationary dynamics.

A further subleading effect relevant for the fibre potential is the higher-derivative $F^4$  correction induced by the supersymmetric completion of the ten-dimensional $\alpha'^3R^4$ sector \cite{Ciupke:2015msa,Cicoli:2016chb,Cicoli:2023njy}: 
\begin{equation}
    V_{F^4}
    =
    -\kappa^2
    \frac{\lambda |W_0|^4}{g_s^{3/2}\mathcal V^4}
    \sum_{i=1}^{h^{1,1}} \Pi_i t^i, \qquad
    \Pi_i
    =
    \int_X c_2(X)\wedge \hat D_i ,
\label{eq:F4-correction}
\end{equation}
where $c_2(X)$ is the second Chern class of the Calabi--Yau threefold and $\hat D_i$ is a basis of harmonic $(1,1)$-forms dual to the divisors $D_i$. The quantities $\Pi_i$ encode the topological dependence on the
Calabi--Yau geometry, while $\lambda$ is an overall numerical
coefficient of the higher-derivative correction. For the nonzero $F^4$ contributions
considered below, we choose $\lambda=-|\lambda|$. These $F^4$ corrections are subleading in the large-volume expansion and can be neglected in the leading LVS stabilisation. However, they may still affect the shape of the fibre potential and will therefore be kept in the inflationary analysis. The consistency requirement that the $F^4$ correction should not spoil
the inflationary plateau gives $|\lambda|\lesssim10^{-6}$ \cite{Cicoli:2017axo}. We therefore adopt
$|\lambda| = \mathcal{O}(10^{-7})$ throughout our analysis.  We also consider the limiting case in which the fibre-dependent
$F^4$ contribution vanishes, denoted by $|\lambda|=0$. Whether this
occurs depends on the compactification data entering the relevant
topological combination; it is not assumed for the specific benchmark
used in Section~\ref{sec:alpha-loop-inflation}. 

The single-field scalar potential used below is therefore organised as
\begin{equation}
    V
    =
    V_{\rm LVS}(\mathcal V,\tau_s)+V_{\rm up}
    +V_{\rm BHP}(\tau_f)+V_{F^4}(\tau_f),
    \label{eq:total-potential-schematic}
\end{equation}
where $V_{\rm up}$ is the positive contribution that raises the LVS AdS
minimum to an approximately Minkowski post-inflationary vacuum, while
the fibre-dependent corrections remain subleading to LVS stabilisation.
Uplift proposals in LVS include warped anti-D3 uplift in
K3-fibred models \cite{Crino:2020qwk,AbdusSalam:2022krp,Cicoli:2024bxw},
dilaton-dependent non-perturbative F-term uplift \cite{Cicoli:2012fh} 
and D-term uplift through coupled F/D-terms
\cite{Burgess:2003ic,Achucarro:2006zf,Cremades:2007ig}. T-brane uplift from positive hidden-matter F-terms was cosidered in \cite{Cicoli:2015ylx,Cicoli:2017shd,Cicoli:2021dhg}, while nilpotent goldstino uplift provide a four-dimensional effective description of the anti-D3 source
\cite{Kallosh:2014wsa,Aparicio:2015psl}, and winding/F-term uplift which create a perturbatively flat direction in the complex structure moduli space by an appropriate flux choice \cite{Gallego:2017dvd,Hebecker:2020ejb,Carta:2021sms,Hebecker:2025tui}. These proposals are subject to mechanism-dependent control of
higher-order and warping corrections
\cite{Ciupke:2015msa,Gao:2022uop,Junghans:2022exo,ValeixoBento:2023nbv}.
For warped anti-D3 uplift, throat and stabilisation fluxes must fit
within the topology-dependent D3 tadpole bound \cite{Gao:2022fdi,Junghans:2022kxg,Bena:2020xrh}. T-brane uplift requires volume, warping and brane-modulus control
\cite{Junghans:2026jza,Junghans:2026hfm}, while the control issue of the F-term uplift was considered in \cite{Hebecker:2025tui}.

We do not claim anti-D3 uplift and alternative uplift mechanisms evade all such constraints. Our focus is perturbative corrections to fibre inflation rather than
the microscopic uplift construction. We assume that a suitable mechanism has uplifted the LVS AdS vacuum
to a metastable, approximately Minkowski vacuum while preserving
the heavy-modulus hierarchy and the BHP channel structure, with no
leading-order modifications to the fibre potential or its slow-roll
derivatives. At fixed heavy-field
values,
\begin{equation}
    V_{\rm const}\equiv
    \left.(V_{\rm LVS}+V_{\rm up})\right|_{
        \mathcal V=\langle\mathcal V\rangle,\,
        \tau_s=\langle\tau_s\rangle}
\end{equation}
enters the single-field potential as a constant; $V_{\rm up}$ need not
be constant before stabilisation. Numerically, we use the leading LVS
stationary values as reference, assuming uplift-induced shifts are
subleading. The constant is fixed by requiring the full fibre potential
to vanish at its post-inflationary minimum. Within these assumptions,
we determine the inflationary predictions and coefficient bounds of
the candidate EFT. We then express the fibre modulus in terms of the
canonically normalised inflaton:
\bea
  \tau_f=\langle\tau_f\rangle e^{k\hat\phi},
    \qquad  k=\frac{2}{\sqrt3}
    \label{eq:canonical}
\eea
up to corrections from the subleading terms. This single-field inflation treatment, driven by \(\hat\phi\) in the fibre direction, is valid only when the heavy LVS moduli remain heavier than the Hubble scale and are not significantly displaced along the inflationary trajectory. This heavy-field analysis is discussed in Section {\ref{sec:BHP-deforme}} and is examined throughout the paper.

One crucial lesson is that the presence of the KK and winding-type BHP terms depends strongly on the brane and
orientifold configuration. 
KK-type corrections arise when there are suitable closed-string KK exchange channels, 
for instance between mutually non-intersecting D7/O7 divisors, with moduli dependence controlled 
by the transverse direction, and generically also from O3--D7/O7 exchange when O3-planes are present. 
Winding-type corrections instead require intersecting D7/O7 sources whose intersection curve has non-trivial one-cycles, 
schematically \(h^{1,0}(D_i\cap D_j)\neq0\).
Consequently, the conventional BHP coefficients can vanish in many
brane configurations, for example when there are no appropriate
intersections or no O3-induced KK exchange channel.

\subsection{Perturbative corrections in type IIB compactifications}

The configuration dependence of the BHP conjecture described above is one of the main reasons to go beyond
the strict parametrisation.  A ten-dimensional effective-field-theory analysis suggests a broader classification of perturbative effects including warping correction, genuine loop correction, as well as local $\alpha'$ correction  higher-dimensional operators \cite{Gao:2022uop}.

First, warping is the classical backreaction of fluxes and localized sources on the internal geometry. In supersymmetric GKP-type backgrounds, the warped K\"ahler potential remains of no-scale type at leading two-derivative order, so warping alone does not generate a potential for the K\"ahler moduli \cite{Giddings:2001yu}. Warping can nevertheless modify K\"ahler coordinates and impose important control conditions, especially in uplifted compactifications \cite{Junghans:2022exo,Gao:2022fdi,Bansal:2024uzr}.

Second, genuine loop corrections arise from finite non-local effects of ten-dimensional or brane-localized fields propagating through the compact space.  They cannot be captured by local operators in 10d or on a brane, and
are analogous to Casimir energy effects. From the four-dimensional perspective they correspond to integrating out KK towers. In an approximately isotropic compactification with four-cycle volume $\tau\sim\mathcal V^{2/3}$,
\begin{equation}
    \delta K_{\rm gl}
    \sim
    \frac{1}{\sqrt{\tau}\,\mathcal V}
    \sim
    \mathcal O(\mathcal V^{-4/3}),
    \qquad
    \delta V_{\rm gl}
    \sim
    \kappa |W_0|^2 g_s\,\mathcal O(\mathcal V^{-10/3}) .
    \label{eq:genuine-loop-scaling}
\end{equation}
This has the same homogeneous volume scaling as the BHP winding contribution, but it is not tied to the same brane-intersection criteria.

Third, local higher-derivative corrections arise from Wilsonian operators in the bulk, on D-branes/O-planes, or on their intersection loci. The leading purely gravitational bulk term is the familiar $\alpha'^3R_{10}^4$ correction. Its reduction linear in the external curvature yields the BBHL correction and its one-loop counterpart \cite{Antoniadis:1997eg,Becker:2002nn},
\begin{equation}
    \delta V_{R_{10}^4}
    \sim
    \kappa |W_0|^2
    (g_s^{-3/2}+g_s^{1/2})\,
    \mathcal O(\mathcal V^{-3}) .
    \label{eq:BBHL-scaling}
\end{equation}
where tree-level part is included in the LVS potential through ~\eqref{eq:LVS-K-W}. The first term of ~\eqref{eq:BBHL-scaling} is also known as the standard $\alpha'$ correction \cite{Balasubramanian:2005zx,Conlon:2005ki, Cicoli:2008va}, which  takes the form 
\begin{equation}
\label{deltaV-leading}
\delta V_{\alpha'} \simeq \kappa \frac{3 \xi |W_0|^2}{4 g_s^{3/2} \mathcal{V}^3}\,,
\end{equation}


Localized operators on D7/O7 systems can generate further corrections. Relevant operators such as a  $R_8^2$ term give corrections to the K\"ahler potential homogeneous of degree ($-1$) in the four-cycle volumes,
\begin{equation}
    \delta K_{R_8^2}\sim g_s f_{-1}(\tau_i),
    \qquad
    f_{-1}(a\tau_i)=a^{-1}f_{-1}(\tau_i) .
    \label{eq:R82-K}
\end{equation}
Their scalar potential contribution starts only at quadratic order because of extended no-scale structure and is therefore parametrically similar to the KK-type BHP term, although the existence and coefficient of the corresponding local operator are not universal:
\begin{equation}
    \delta V_{R_8^2}
    \sim
    \kappa |W_0|^2 \,
      g_s^2\,
    \mathcal{O}\!\left(\frac{1}{\mathcal{V}^{10/3}}\right).
    \label{eq:deltaV-R82}
\end{equation}
By contrast, marginal localized operators, such as a possible $R_8^4$ operator on a D7/O7 worldvolume or an $R_6^3$ operator on an intersection locus, can generate logarithmic degree ($-2$) corrections,
\begin{equation}
    \delta K_{\log}
    \sim
    \ln(M_{10}g_s^{1/4}L)\,f_{-2}(\tau_i),
    \qquad
    f_{-2}(a\tau_i)=a^{-2}f_{-2}(\tau_i) .
    \label{eq:log-local-K}
\end{equation}
These terms enter the scalar potential at the same inverse-radius order as genuine loops, but their microscopic origin is local rather than non-local:
\begin{equation}
    \delta V_{\rm log}
    \sim
    \kappa |W_0|^2\,    \ln\!\left(M_{10}g_s^{1/4} \cV^{{1}/{6}}\right)
    \mathcal{O}\!\left(\frac{1}{\mathcal{V}^{10/3}}\right).
    \label{eq:deltaV-marginal-revised}
\end{equation}

The important lesson is that the fibre potential is not determined by
the Calabi--Yau volume form alone. It also depends on the orientifold
projection,  the specific D7/O7/O3 configuration and the wrapped divisors. The conventional BHP
coefficients may be present, partially absent, or vanish altogether.
At the same time, the effective-field-theory analysis of fibred
geometries suggests additional degree $(-2)$ in four-cycle volume fibre-dependent corrections
with the same large-volume order as winding-type effects, including
genuine non-local loop contributions and possible logarithmic terms
from marginal localized operators. We now specialize to this
fibred-geometry sector and derive the effective correction used in the
inflationary analysis.

\section{Constraints on \texorpdfstring{$\alpha^\prime$}{alpha-prime}-loop corrections in fibre inflation}
\label{sec:BHP-deforme}

We now specialize the general discussion of Section~\ref{sec:bg} to
fibred geometries and study how the generalized \(\alpha'\)-loop sector
affects fibre inflation. The coefficients of this sector are not
computed from a microscopic Calabi--Yau metric or explicit string
amplitudes. Instead, they are treated as effective parameters encoding
possible genuine loop effects, D7/O7-sector threshold contributions and
marginal higher-derivative data with the same fibre-dependent moduli
structures.

A central point of this section is the following crucial distinction:
the same \(\alpha'\)-loop coefficient can represent either a small
deformation of an existing fibre potential or a leading inflationary
contribution, depending on the conventional BHP baseline and the
available KK- and winding-type channels in the underlying orientifold
and D-brane configuration. We therefore distinguish three classes of BHP baselines:
\begin{itemize}
    \item
    \textbf{BHP-complete configurations}, where both KK- and winding-type BHP structures contribute to the fibre potential. In this case the
    \(\alpha'\)-loop sector is treated as an additional deformation of
    the standard fibre-inflation potential. The corresponding
    constraints are obtained by requiring that the deformation does not
    spoil the inflationary plateau.

    \item
    \textbf{Partially BHP-driven configurations}, where only one of the
    conventional BHP channels is present. This includes KK-only and
    winding-only baselines. In these cases, the \(\alpha'\)-loop
    corrections associated with the missing BHP structure need not be
    small and can provide an essential contribution to the fibre
    potential. Their absolute sizes are therefore not directly comparable to the fractional deformations in BHP-complete configurations.

    \item
    \textbf{BHP-free configurations}, where both conventional KK- and
    winding-type BHP channels vanish for the chosen brane/orientifold
    configuration. The generalized \(\alpha'\)-loop sector can then
    become the leading source of the fibre potential and support
    inflation, which we refer to as ``\(\alpha'\)-loop inflation''.
    This regime is analysed separately in
    Section~\ref{sec:alpha-loop-inflation}.
\end{itemize}

The first two classes are analysed in this section, using explicit
K3-fibred Calabi--Yau orientifold examples to determine the allowed
deformations and viable benchmark regions. The BHP-free regime is
studied separately in Section~\ref{sec:alpha-loop-inflation}, where the
generalized \(\alpha'\)-loop sector is treated as the leading source of
the inflaton potential.

\subsection{Effective \texorpdfstring{$\alpha'$}{alpha-prime}-loop corrections in fibred geometry}
\label{sec:alpha-loop-correction}

Following the two-step effective-field-theory analysis of
~\cite{Gao:2022uop}, we now focus on the subset of corrections
which are relevant for the fibre direction in a K3-fibred
compactification. This is not meant to be a complete list of all
perturbative corrections in Type IIB orientifolds. 
Rather, it isolates the minimal fibre-dependent degree \((-2)\)
sector in four-cycle volume variables suggested by large-volume power counting
in the hierarchical regime
\(\tau_b\gg\tau_f>\tau_s\) as in (\ref{eq:fibred-volume-review}).

For the K3-fibred geometries relevant to fibre inflation, the EFT
power-counting analysis suggests a collection of candidate
fibre-dependent structures of schematic form ~\cite{Gao:2022uop}
\begin{equation}
\begin{split}
    \delta K_{\rm cand}
    \sim &\;
    \frac{1+\ln\tau_f+\ln\tau_b}{\tau_b^2}
    +
    \frac{1+\ln\tau_f}{\tau_b\tau_f}
    +
    \frac{1+\ln\tau_f+\ln\tau_b}{\tau_f^2}
    \\
    &\quad
    +
    \frac{\tau_f(1+\ln\tau_f)}{\tau_b^3}
    +
    \frac{1}{\mathcal V\sqrt{\tau_s}}
    +
    g_s f_{-1}(\tau_i)
    +\cdots ,
\end{split}
\label{eq:deltaK-candidate-full}
\end{equation}
where the origin of these terms  is reviewed in Appendix~\ref{app:alpha_loop_origin}.
 The last
term represents degree $(-1)$ in four-cycle volume local effects, such as 
$R_8^2$ operator on D7/O7 systems or the corresponding intersection terms.
Because of the extended no-scale cancellation, such terms contribute to
the scalar potential only at quadratic order in $\delta K$ and are more
naturally grouped with the KK-type BHP or local-KK sector, rather than
with the degree $(-2)$ fibre ansatz studied here.  The blow-up term
$1/(\mathcal V\sqrt{\tau_s})$ affects the small-cycle sector and is not
included in the single-field fibre potential.  The term proportional to
$\tau_f/\tau_b^3$ is associated with an inverse-fibration scaling and is
subleading in the hierarchy \eqref{eq:fibred-volume-review}.  Purely
base-dependent terms are also suppressed once the large volume is fixed.
Finally, logarithms multiplying the pure $1/\tau_f^2$ or $1/\tau_b^2$
structures would introduce additional independent functions in the
kinetic terms. Since their survival depends on tensor contractions,
field redefinitions or cancellations which are not known in a generic
Calabi--Yau orientifold, we omit them and retain the following minimal ansatz in the fibred-geometry
sector (see Appendix~\ref{app:alpha_loop_origin}):
\begin{equation}
    \delta K_{\alpha'\text{-loop}}
    \simeq
    \frac{C_1}{\tau_b^2}
    +
    \frac{C_2}{\tau_f^2}
    +
    \frac{
        C_3+C_4\ln \tau_f
    }{\tau_b\tau_f}.
    \label{eq:deltaK-alpha-loop-fibred}
\end{equation}
Here ``$\alpha'$-loop'' is shorthand for this restricted class of corrections; it neither exhausts all possible local $\alpha'$ and loop effects nor implies that the terms in \eqref{eq:deltaK-alpha-loop-fibred} originate from a single universal operator.
The coefficients in \eqref{eq:deltaK-alpha-loop-fibred} have different
microscopic interpretations and moduli dependence, which will be
constrained phenomenologically since exact values in \eqref{eq:deltaK-alpha-loop-fibred} would
require knowledge of the Calabi-Yau metric, the orientifold projection,
the D7/O7 configuration and the relevant string amplitudes. 
\begin{itemize}
    \item
    $C_2$ is the  degree $(-2)$ in four-cycle volume coefficient with
    fibre-scale dependence.  A natural source is the genuine non-local
    loop effect obtained by first compactifying the ten-dimensional
    theory on the fibre direction, integrating out the corresponding heavy KK modes of the fibre, and then compactifying to four dimensions. Additional brane-localized genuine loop effects can also
    be absorbed into $C_2$ whenever their moduli dependence reduces to
    the same $1/\tau_f^2$ structure.

    \item
    \(C_3\) parametrises the finite non-logarithmic mixed
degree \((-2)\) structure, which may receive contributions from
D7/O7-localized operators and genuine loop effects with the same
moduli dependence. It should not be identified with the
    relevant $R_8^2$ operator alone, the latter gives a degree $(-1)$
    K\"ahler potential correction and belongs to the KK-like BHP sector.

    \item
    $C_4$ controls the logarithmic mixed contribution.  It is associated
    with the logarithmic running of candidate marginal localized
    operators, such as an $R_8^4$ operator on a D7/O7 worldvolume or an
    $R_6^3$ operator on a D7/O7 intersection locus.  Its existence,
    tensor structure and coefficient are model dependent.

    \item
    $C_1$ denotes the analogous base-sector degree $(-2)$ contribution.
    It will be dropped from
    the single-field fibre potential because $\tau_b\gg\tau_f$ in the LVS
    regime.
\end{itemize}

Using the linearized no-scale formula (\ref{eq:linear-deltaV})
and the fact that all terms in
\eqref{eq:deltaK-alpha-loop-fibred} are homogeneous of degree $(-2)$
up to the logarithm, we find
\begin{equation}
    V_{\alpha'\text{-loop}}
    \simeq
    -\kappa
    \frac{|W_0|^2}{\mathcal V^2}
    \left[
        \frac{2C_2}{\tau_f^2}
        +
        \frac{
            2C_3+
            2 C_4\left(\ln \tau_f -\frac{3}{2}\right)
        }{\tau_b\tau_f}
    \right].
    \label{eq:alpha-loop-potential-C}
\end{equation}

In the following,  we use the fibre-volume normalization with scaling parameter $\alpha$
\begin{equation}
    \mathcal V \simeq \frac{1}{\sqrt{2\alpha}}\sqrt{\tau_f}\,\tau_b,
    \qquad
    \frac{1}{\tau_b\tau_f}
    =
    \frac{1}{\sqrt{2\alpha}\,\mathcal V\sqrt{\tau_f}},
    \label{eq:fibred-volume-normalization}
\end{equation}
and defining
\begin{equation}
    \theta_1=-2C_2,
    \qquad
    \theta_2=\frac{2C_3}{\sqrt{2\alpha}},
    \qquad
    \theta_3=\frac{2C_4}{\sqrt{2\alpha}},
    \label{eq:theta-definitions}
\end{equation}
the fibre-dependent correction leads to the so-called  ``$\alpha'$-loop" potential:
\begin{equation}
    V_{\rm \alpha'-loop}
    =
    \kappa\frac{|W_0|^2}{\mathcal V^2}
    \left[
        \frac{\theta_1}{\tau_f^2}
        -
        \frac{\theta_2+\theta_3\left(\ln \tau_f-\frac32\right)}
             {\mathcal V\sqrt{\tau_f}}
    \right].
    \label{eq:alpha-loop-potential}
\end{equation}

When conventional BHP terms are present as in $V_{\rm BHP}$  (\ref{eq:BHP-potential}), the non-logarithmic
\(\theta_1\) and \(\theta_2\) contributions project onto the same
single-field fibre structures as the corresponding BHP coefficients,
schematically  $A_{\rm eff}=A+\theta_1$ and
    $B_{\rm eff}=B+\theta_2$. This is not
a microscopic identification of the two effects, their higher-dimensional
origins and selection rules may be different. Rather, it means that after
reducing to the four-dimensional fibre potential they renormalise the
same power-law operators, while the logarithmic \(\theta_3\) term remains
a genuinely new functional deformation. The scan therefore constrains the
allowed size of additional EFT contributions around a chosen BHP
baseline.

A particularly interesting question arises when BHP coefficients
vanish because the required transverse or intersection channels are absent: can the $\alpha'$-loop corrections alone support a viable inflationary phase? 
The central point  is that the BHP selection rules do
not exhaust the possible degree $(-2)$ corrections in the EFT.  Genuine
non-local loops have the same homogeneous degree as winding-type BHP terms
but need not originate from intersecting D7/O7 sources.  Candidate marginal
localized operators can also generate logarithmic degree $(-2)$ terms.  In
this sense the $\theta_i$ sector can survive in configurations in which the
conventional BHP coefficients vanish while
the generalized degree $(-2)$ effects in \eqref{eq:alpha-loop-potential}
can nevertheless survive.  
This observation motivates the study of BHP-free fibre inflation in Section \ref{sec:alpha-loop-inflation}.

\subsection{Inflationary consistency conditions and scan strategy}
\label{sec:bhp-alpha-loop-f4}

For a K3-fibred Swiss-cheese volume,  the fibre-dependent scalar potential can receive corrections from  conventional BHP loop corrections (\ref{eq:BHP-potential}), the higher-derivative $F^4$ correction (\ref{eq:F4-correction}), and the fibred $\alpha'$-loop corrections (\ref{eq:alpha-loop-potential}):
\bea
V_{\rm corr}(\tau_f)
=
V_{BHP}^{\rm W}(\tau_f)
+
V_{BHP}^{\rm KK}(\tau_f)
+
V_{F^4}(\tau_f)
+
V_{\alpha^\prime{\rm -loop}}(\tau_f),
\label{eq:inf-potential}
\eea
where the terms that are absent for a given D7/O7/O3 configuration are simply omitted. This unified form is useful since the same analysis can be applied to BHP-complete, partially BHP-driven and BHP-free configurations simply by removing the absent channels from \(V_{\rm corr}\). 

After the LVS heavy fields and the blow-up modulus $\tau_s$ are fixed at their leading values, the fibre modulus is expressed in terms of the canonically normalised inflaton \(\hat\phi\) using \eqref{eq:canonical}: $\tau_f=\langle\tau_f\rangle e^{\frac{2}{\sqrt3}\hat\phi}$.
The full single-field  potential is then
\begin{equation}
    V_{\rm inf}(\hat\phi) = V_{\rm const} +V_{\rm corr}
    =
    \frac{\kappa |W_0|^2}{\langle \cV \rangle^2}C_{\rm dS}
    +
    V_{\rm corr}\!\left(\langle\tau_f\rangle e^{k\hat\phi}\right),
\end{equation}
Here \(C_{\rm dS}\) parametrises \(V_{\rm const}\) in Eq.~(2.14);
the approximately Minkowski post-inflationary condition fixes its
required value within the assumed uplifted EFT. The potential vanishes at the post-inflationary  minimum \(\hat\phi=0\), or equivalently \(\tau_f=\langle\tau_{f}\rangle\). The consistency of this single-field treatment requires the heavy  modulus to remain sufficiently massive and weakly displaced along the trajectory.

The structure of the single-field potential makes clear why the bounds on \(\alpha'\)-loop coefficients are configuration dependent. If both BHP terms are present, the \(\theta_1\) and \(\theta_2\) terms project onto the same single-field power-law structures as deformations of the corresponding BHP coefficients. By contrast, the \(\theta_3\) term is a genuine logarithmic deformation of the single-field potential and can affect both the slope and the curvature. The purpose of the following scans is to determine how large these deformations can be before the fibre-inflation plateau is spoiled.


We first fix the orientifold Calabi-Yau threefolds $X_3$, together with the brane configuration and LVS-stabilisation parameters to the benchmark values quoted below. We then scan the \(\alpha'\)-loop coefficients $(\theta_1,\theta_2,\theta_3)$. 
The parameter $\lambda$ is either fixed to the benchmark value of the corresponding setup, or set to zero when the $F^4$ correction is switched off. 
 We mainly display the sign choices relevant for positive deformations of the BHP-driven potentials, i.e., $\theta_1\ge 0$, $\theta_2\le 0$ and $\theta_3\le 0$,  but other signs were also checked and do not affect the qualitative conclusions. 

 Several constraints should be imposed simultaneously in the analysis:
\begin{enumerate}
\item  {\bf Cosmological validity control:}
For every scan point  we compute 
\begin{equation}
    \epsilon_V=\frac{1}{2V_{\rm inf}^2}\left(\frac{\partial V_{\rm inf}}{\partial \hat\phi}\right)^2,
    \qquad
    \eta_V=\frac{1}{V_{\rm inf}}\frac{\partial^2 V_{\rm inf}}{\partial \hat\phi^2}.
\end{equation}
The end of inflation is determined by \(\epsilon_V(\hat\phi_{\rm end})=1\), and the horizon-exit point  $\hat\phi_\ast$ is obtained from the scalar-amplitude condition \cite{Planck:2018jri, Planck:2018vyg,Kallosh:2025ijd}
\begin{equation} \label{power_spectrum}
    \mathcal P_{\mathcal R}
    =
    \left.\frac{V_{\rm inf}}{24\pi^2\epsilon_V}\right|_{\hat\phi=\hat\phi_\ast}
    =2.1\times 10^{-9}\,.
\end{equation}
The number of e-folds and the inflationary observables are then evaluated as
\begin{equation}
    N_e=\int_{\hat\phi_{\rm end}}^{\hat\phi_\ast}\frac{V_{\rm inf}}{V'_{\rm inf}}\,d\hat\phi,
    \qquad
    r=16\epsilon_V,
    \qquad
    n_s=1-6\epsilon_V+2\eta_V\,,
    \label{eq:observation}
\end{equation}
and we impose the constraints from the Planck 2018 \cite{Planck:2018jri} and the BK18 \cite{BICEP:2021xfz}:
\begin{equation}
\label{eq:cosmological}
    45\lesssim N_*\lesssim65,
    \qquad
    n_s=0.9649\pm0.0042,
    \qquad
    r_{0.05}<0.036.
\end{equation}
where the adopted interval for \(N_*\) represents a deliberately broad
reheating uncertainty.

\item {\bf Leading expansion control:} We  require the large-volume expansion parameter to be small,
\begin{equation}
    \epsilon_{\alpha^\prime}
    \equiv
    \frac{\xi}{2g_s^{3/2}\mathcal V}
    \ll 1 .
    \label{eq:epsilon}
\end{equation}
We additionally impose
\bea
    \left|  V_{\rm corr}(\tau_f) \right|/ \left| V_{\rm LVS}(\tau_s,\mathcal V) \right| < \cO(0.1)
  \label{eq:ratio}  
\eea
as a conservative preliminary screening criterion. This condition
requires the fibre-dependent sector not to parametrically overwhelm the
leading LVS stabilisation potential. It does not by itself guarantee
heavy-field decoupling, which is tested independently below.

\item {\bf Deformation control:  }  An additional condition is imposed for BHP-baseline  inflationary scenarios and should not be applied to $\alpha'$-loop dominated regime such as the KK-only and BHP-free cases. Let $\langle\tau_{f,0}\rangle$ denote the fibre minimum obtained in the absence of the $\alpha^\prime$-loop corrections, and let $\langle\tau_f\rangle$ denote the corresponding minimum after including them. We impose
\begin{equation}
    \frac{
        \left|\langle\tau_f\rangle-\langle\tau_{f,0}\rangle\right|
    }{
        \langle\tau_{f,0}\rangle
    }
    <0.05 ,
    \label{eq:deforme}
\end{equation}
such that the correction does not spoil the BHP vacuum.

\item {\bf Heavy moduli control:} The volume modulus should remain sufficiently heavier than the inflationary Hubble scale. We therefore require
\bea
    \delta_{\mathcal V}
    \equiv
    \frac{H_*}{m_{\mathcal V}}
    \simeq
    \sqrt{
    \frac{V_{\rm inf}(\hat{\phi}_*)}
    {3\rho_{\mathcal V}V_{\alpha'^3}}
    }
    <1 ,  
\label{eq:heavy-modulus-hierarchy}
\eea
where \begin{equation}
    m_{\cV}^2
    =
    \rho_{\cV}\,V_{\alpha^{\prime 3}},
    \qquad
    V_{\alpha^{\prime 3}}
    =
    \kappa\frac{|W_0|^2}{\cV^3}
    \frac{3\xi}{4g_s^{3/2}},
    \label{eq:rhoV-def}
\end{equation}
and $\rho_{\mathcal V}$ is a model-dependent coefficient determined by
the covariant Hessian of the heavy sector, including the effects of the
uplift and the $\tau_s$--$\mathcal V$ mixing. The standard
order-of-magnitude estimate $m_{\mathcal V}^2\sim V_{\alpha'^3}$,
corresponding to $\rho_{\mathcal V}=\mathcal O(1)$, is commonly used in
fibre-inflation analyses; see, for example, Eq.~(4.10) of~\cite{Cicoli:2017axo}. We retain $\rho_{\mathcal V}$ explicitly in
the analytical expressions and adopt the fiducial choice
$\rho_{\mathcal V}=1$ in the numerical estimates. For the fiducial choice \(\rho_{\mathcal V}=1\), this condition provides an order-of-magnitude check of the single-field hierarchy
within the assumed uplifted EFT. For  conservativeness, we adopt the stronger condition along the inflationary trajectory as:
\bea
    \max_{\rm traj}\,\, \delta_{\mathcal V} < 0.1 
\eea
Independently, the heavy-field displacements must satisfy
\begin{equation}
\frac{\delta \cV}{\cV} \ll 1,\qquad \frac{\delta \tau_s}{\tau_s} \ll 1 \label{eq:heavy-field-shift}
\end{equation}

\item {\bf K\"ahler cone control:} The full slow-roll trajectory must remain inside the K\"ahler cone. This prevents singularities from appearing in the inflationary potential and ensures the validity of the geometric effective-field-theory description. Thus, for every scan point we impose
\begin{equation}
    \hat\phi_{\rm KC}^{\rm min}
    <
    \hat\phi_{\rm end}
    <
    \hat\phi_\ast
    <
    \hat\phi_{\rm KC}^{\rm max}.
\end{equation}
The values of $\hat\phi_{\rm end}$ and $\hat\phi_*$ are determined
from the full potential at each scan point, and the complete trajectory
is required to remain inside the K\"ahler cone.

\end{enumerate}

In the following,   we employ two complementary scan procedures.
First, we perform a {\it one-parameter scan} of the
\(\alpha'\)-loop sector. For each scan, the geometric, LVS, BHP and
\(F^4\) background is held fixed, one coefficient \(\theta_i\) is
varied, and the other two are set to zero:
\begin{equation}
    \theta_i\neq0,
    \qquad
    \theta_j=0
    \quad (j\neq i).
\end{equation}
For every trial value, we recompute the fibre minimum, the inflationary
trajectory and all consistency conditions described above.
These one-dimensional scans measure  the order of magnitude at which an
isolated \(\theta_i\) contribution begins to spoil the reference
potential. The resulting intervals are therefore upper bounds on
single-coefficient deformations of the chosen baseline. They are not
component-wise bounds on configurations in which several coefficients
are simultaneously nonzero.

Second, we perform simultaneous {\it multi-parameter scans}, with the
strategy depending on the BHP baseline. For BHP-complete and winding-only configurations, a viable
fibre-inflation background already exists before introducing the
generalized \(\alpha'\)-loop sector. We therefore fix the corresponding
LVS, BHP and \(F^4\) benchmark parameters and scan the additional 
coefficients simultaneously
\bea
\label{eq:smallscan}
\theta_i=\{0,\pm10^j\},\quad j\in[-7,7].
\eea
The scan is performed on a logarithmic grid in the magnitudes of the
parameters.
At each point, the fibre minimum, inflationary
trajectory and all consistency conditions are recomputed. This scan
captures correlations and partial cancellations among the deformation
terms and can therefore allow individual coefficients outside the
one-parameter envelopes.

For KK-only and BHP-free configurations, in contrast, the generalized
\(\alpha'\)-loop sector can provide structures absent in the original
BHP potential. We therefore perform broader scans over the effective
parameter space, including LVS parameters $
\{W_0,A_s,g_s\}$, non-vanishing BHP parameters $C_{\rm BHP}^{(a)}$, $F^4$ correction parameter $\lambda$ and $\alpha'$-loop parameters $\theta_i$:
\bea
\label{eq:parameters}
\cP_{\rm scan}=(W_0,A_s,g_s, C_{\rm BHP}^{(a)}, \lambda,\theta_1,\theta_2,\theta_3),
\eea
with
\begin{align}
&0.1<W_0<500,\qquad
1<A_s<100,\qquad
0.05<g_s<0.5, \qquad |C_{\rm BHP}^{(a)}| < 1000, \nonumber\\
&\lambda=0\quad\text{or}\quad
10^{-7}\leq|\lambda|\leq10^{-3},
\qquad
\theta_i=\{0,\pm10^j\},\quad j\in[-7,7]. \label{eq:parameter-range}
\end{align}
The Calabi--Yau geometry, orientifold involution and brane configuration
are kept fixed. For every trial point, the LVS minimum, fibre minimum,
inflationary trajectory and heavy-field stability conditions are
recomputed.
Therefore, the role of the simultaneous scan is different in the two
cases: it characterises correlated deformations of an existing BHP
inflationary background in BHP-driven branches, while it searches for
new viable inflationary regimes in KK-only and BHP-free branches.

\subsection{\texorpdfstring{$\alpha'$}{alpha-prime}-loop corrections in BHP baselines}
The Calabi--Yau geometries, orientifold involutions and D7/O7 configurations used below are taken from \cite{Cicoli:2016xae,Cicoli:2017axo}, where conventional fibre-inflation models were analysed in orientifold Calabi--Yau compactifications. We do not repeat this global analysis here. Instead, we use these compactifications as fixed benchmark backgrounds and ask how the additional fibred-geometry $\alpha^\prime$-loop deformations modify the corresponding fibre potentials.

\subsubsection{\texorpdfstring{Example 1:  Calabi--Yau threefold with $h^{1,1} = 3$}{Example 1: Calabi--Yau threefold with h11 = 3}}
\label{sec:example1-kk-winding-alpha-loop-f4}

Our first benchmark is the K3-fibred Calabi--Yau threefold $X_3$ with Hodge numbers $(h^{2,1},h^{1,1})=(99,3)$, defined by the GLSM charge matrix \cite{Cicoli:2016xae}:
\begin{table}[H]
  \centering
 \begin{tabular}{|c|ccccccc|}
\hline
     & $x_1$  & $x_2$  & $x_3$  & $x_4$  & $x_5$ & $x_6$  & $x_7$       \\
    \hline
6 & 0  & 0 & 1 & 1 & 1 & 0  & 3   \\
8 & 0  & 1 & 1 & 1 & 0 & 1  & 4   \\
8 & 1  & 0 & 1 & 0 & 1 & 1  & 4   \\   \hline
  & dP$_8$  & NdP$_{10}$ & SD$_1$ &  NdP$_{15}$ & NdP$_{13}$ & K3  &  SD$_2$  \\
    \hline
  \end{tabular}
 \end{table}
 with the Stanley--Reisner (SR) ideal 
 \begin{equation}
    \mathrm{SR}
    =
    \left\{
    x_1x_5,\,
    x_1x_6x_7,\,
    x_2x_3x_4,\,
    x_2x_6x_7,\,
    x_3x_4x_5
    \right\}.
\end{equation}
The last row of the table indicates the topology of each divisor $D_i\equiv \{x_i = 0\}$, where $D_6$ is the K3 surface, $D_{1,2,4,5}$ are $dP_{8,10,15,13}$ surfaces and $D_{3,7}$ are the special deformation surfaces defined in \cite{Gao:2013pra}. The corresponding four-cycle volumes are denoted by $\tau_i$.
In the basis adapted to the fibre direction, $\tau_f = \tau_7 - 2\tau_6 + 3\tau_1$, $\tau_s = \tau_1$ and $\tau_b = \tau_6$, the relevant volume form is
\bea
   \cV
    =
    \frac{1}{6}\sqrt{\tau_f}\,\tau_b
    -
    \frac{\sqrt2}{3}\tau_s^{3/2}.
    \label{eq:sec32-example1-volume}
\eea
where the fibre modulus is \(\tau_f\), \(\tau_b\) controls the base direction and \(\tau_s\) is the blow-up modulus.

The orientifold structures of this geometry were also analysed in the database \cite{Altman:2021pyc, Crino:2022zjk, Cao:2024oqx}
\footnote{In determining the fixed loci of involutions, the
fixed-point conditions should ultimately be solved modulo the full
complex toric scaling group $(\mathbb C^\ast)^r$.  For
single-coordinate reflections, the fixed loci obtained from a
real-sector search reproduce the corresponding results of
~\cite{Crino:2022zjk}.  Such a restriction, however, need not
exhaust all solutions and ~\cite{Cao:2024oqx} extends the analysis
to the full complex toric scaling space, where additional fixed loci
can occur.}.
For this geometry, it was shown in \cite{Cao:2024oqx} that there are no non-trivial identical-divisor (NID) exchange involutions, while various reflection (REF) involutions can be performed.
We consider the brane configurations under two involutions in  \cite{Cicoli:2016xae}:
\begin{itemize}
    \item BHP-complete case, $x_3 \leftrightarrow -x_3$: the fixed locus consists of one O7-plane wrapping $D_3$, three O3-planes on $D_1D_2D_7$ and two O3-planes on $D_4D_5D_7$. Two stacks of D7-branes wrap $D_2$ and $D_5$ to cancel the D7 tadpole, and D3-tadpole cancellation is satisfied for the total D3 charge $|Q_{D3}| = 9$. Winding-type corrections arise from the intersections of the D7-branes with the O7-plane on $D_2\cap D_3$, or from the D7-brane intersection on $D_2\cap D_5$, because $h^{1,0}(D_2\cap D_{3,5})\neq0$. Since all D7-branes intersect the O7-plane, the KK-type corrections arise only from KK string loops between the O3-planes and D7/O7 sources.
    \item KK-only BHP case, $x_6 \leftrightarrow -x_6$: the fixed locus consists of one O7-plane on $D_6$, six O3-planes on $D_1D_4D_7$ and six O3-planes on $D_2D_5D_7$. The D7-branes lie on top of the O7-plane, while the D-branes and O-planes do not intersect. Thus, the winding contribution is absent and only a KK-type BHP term from KK string loops between the O3-planes and D7/O7 sources is present.
\end{itemize}

\subsubsection*{ BHP-complete baseline}

For the brane setup with both BHP channels, 
 the conventional fibre potential $V_{BHP}$ can be written following (\ref{eq:BHP-potential}) with
\begin{align}
    A &= \frac{1}{4} \left(g_s\,C_f^{KK}\right)^2, \qquad
     B =  C_1^W - \frac{C_2^W}{1 - \sqrt{\frac{\tau_s}{2 \tau_f}}}, \nonumber \\
    C &= \frac{(g_s C_b^{KK})^2}{72} \big[ 1 - 6 \frac{C_s^{KK}}{C_b^{KK}} \sqrt{\frac{2 \tau_s}{\tau_f}} + \frac{{ C_f^{KK}}}{C_b^{KK}} (\frac{2 \tau_s}{\tau_f})^{3/2} \big] \,,
    \label{eq:11ABC}
\end{align}
where $C_1^{\rm W}$ and $C_2^{\rm W}$ denote the winding-loop coefficients associated with the intersection curves $D_2\cap D_3$ and $D_2\cap D_5$, respectively. Similarly, $C_f^{\rm KK}$, $C_b^{\rm KK}$ and $C_s^{\rm KK}$ are the KK-loop coefficients associated with the fibre, base and blow-up directions. These constants encode the dependence on the complex-structure moduli and on the detailed D7/O7-brane configuration.

The $F^4$ corrections~\eqref{eq:F4-correction} take the special form $V_{F^4}$:
\begin{equation} \label{eq:f4-example1}
    V_{F^4} = \kappa \frac{F W_0^2}{\mathcal{V}^2} \left(\frac{2}{3\mathcal{V} \tau_f} + \frac{\sqrt{\tau_f}}{\mathcal{V}^2} \right) ,
\qquad {\rm with} \qquad
    F = 36|\lambda|\kappa W_0^2 g_s^{-3/2}\,.
\end{equation} 
Including the $\alpha'$-loop sector $V_{\rm \alpha'-loop}$ in \eqref{eq:inf-potential}, the inflationary potential takes the form:
\begin{align}
\label{inf_1} 
    V_{\text{inf}} =& \kappa \frac{W_0^2}{\mathcal{V}^2} \bigg\{\mathcal{C}_{\text{dS}} + \frac{A + \theta_1}{\langle\tau_f\rangle^2}e^{-2k\hat{\phi}} - \frac{\theta_3 (\ln\langle\tau_f\rangle + k\hat{\phi} - 3/2)}{\mathcal{V}\sqrt{\langle\tau_f\rangle}}e^{-k\hat{\phi}/2} - \frac{B + \theta_2}{\mathcal{V} \sqrt{\langle\tau_f\rangle}}e^{-k\hat{\phi}/2} \nonumber \\
    &+ \frac{2F}{3\mathcal{V}\left<\tau_f\right>}e^{-k\hat{\phi}} + \frac{C \langle\tau_f\rangle}{\mathcal{V}^2}e^{k\hat{\phi}} + \frac{F \sqrt{\langle\tau_f\rangle}}{\mathcal{V}^2}e^{k\hat{\phi}/2} \bigg\},
\end{align}
with the  normalisation factor $k= \frac{2}{\sqrt{3}}$. 
Here \(\theta_1\) and \(\theta_2\) act as effective deformations of the BHP coefficients $A$ and $B$, while \(\theta_3\) is an additional logarithmic deformation of $B$.

The K\"ahler cone of the explicit Calabi--Yau geometry $X_3$ was
computed in~\cite{Cicoli:2016xae}. After fixing the heavy
directions at their stabilised values, its restriction to the
fibre-inflation trajectory becomes
\begin{align}\label{eq:kahler-cone-tau-1}
   2\left<\tau_s\right> < \tau_f < (3\mathcal{V})^{2/3},
\end{align}
and, in terms of the inflaton $\hat{\phi}$, the inflaton range is constrained by
\begin{align}\label{eq:kahler-cone-phi-1}
    \frac{\sqrt{3}}{2}\ln{\left( \frac{2\left<\tau_s\right>}{\left<\tau_f\right>} \right)} < \hat{\phi}_{\rm end} <\hat{\phi} < \hat{\phi}_* < \frac{\sqrt{3}}{2}\ln{\left( \frac{(3\mathcal{V})^{2/3}}{\left<\tau_f\right>} \right)}.
\end{align} 

In the following, we use the benchmark values\footnote{We do not use the benchmark values of \cite{Cicoli:2016xae}, since they do not satisfy the latest observational constraints.}:
\begin{align}
    C_1^W &= -2.6564 ,\,\quad C_2^W = -2.4927 ,\,\quad C_f^{KK} = 0.169 ,\,\quad C_b^{KK} = -2.205 ,\nonumber \\ C_s^{KK} &= -0.16 ,\,\quad W_0 = 347 ,\,\quad g_s = 0.4894 ,\,\quad a_s = 2 \pi ,\, \quad A_s = 3.2 \,
    \label{eq:bench1}
\end{align}
which gives the inflationary potential  \eqref{inf_1} valued at the leading LVS minimum, given by
\begin{align}
    \left<\tau_s\right> \simeq 1.33, \,\quad \mathcal{V} \simeq 27619 \,.
\end{align}
At this benchmark point, we realise the standard BHP-type fibre-inflation scenario shown in Table~\ref{tab:BHP-complete}, with heavy-modulus control parameter $\delta_\cV \simeq 0.05$, inflationary energy scale
$V_{\rm inf}^{1/4} \sim (8 \times 10^{-13})^{1/4}M_p \simeq 2.3\times10^{15}\,\mathrm{GeV}$ and expansion-control parameter $\epsilon_{\alpha'} \simeq 2.5 \times 10^{-5}$.
\begin{table}[ht]
\small
\centering
\begin{tabular}{c|cccccccc}
\toprule
Scenario & $|\lambda|$ & $\langle\tau_{f,0}\rangle$ & $N_e$ & $\hat\phi_{\rm end}$ & $\hat\phi_*$ & $n_s$ & $r$ & $V_{\rm corr}/V_{\rm LVS}$ \\
\midrule
BHP & 0 & 50.62 & 63.1 & 0.776 & 2.58 & 0.96793 & $2.7 \times 10^{-5}$ & 0.0647\\
BHP + $F^4$ & $1 \times 10^{-7}$ & 51.25 & 60.7  & 0.775 & 2.56 & 0.9629 & $2.6 \times 10^{-5}$  & 0.0593
 \\
\bottomrule
\end{tabular}
\caption{BHP-complete baseline in Example~1.}
\label{tab:BHP-complete}
\end{table}

\begin{figure}[ht]
	\centering 
\includegraphics[width=0.47\textwidth]{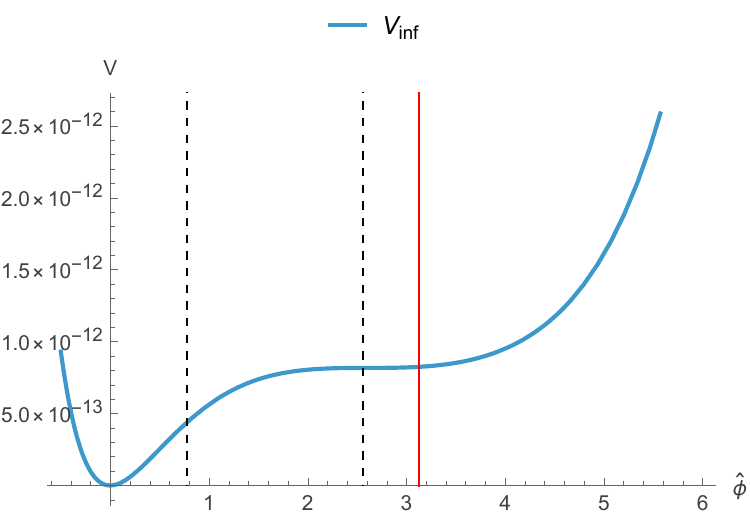}
\hfill
\includegraphics[width=0.47\textwidth]{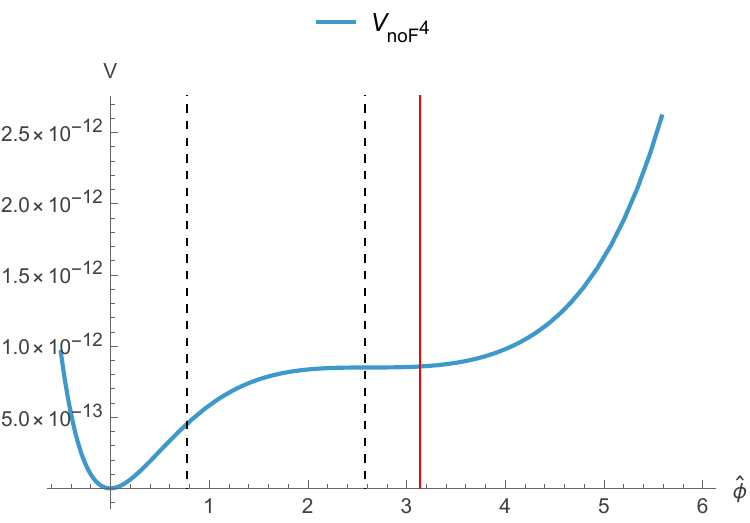} 
	\caption{
Inflationary potentials in Example~1 with and without the $F^4$ correction. 
}
	\label{fig:V_inf}
\end{figure}

Now, we turn on the  $\alpha^\prime$-loop corrections, characterized by the coefficients $\theta_i$,  and examine their effects on the BHP-complete  fibre potential.  The resulting inflationary potentials are shown in Figure~\ref{fig:V_inf}, where the red solid vertical line denotes the K\"ahler cone bound, and the black dashed lines indicate horizon exit and the end of inflation. 

The results of the one-parameter scan are summarised in Table~\ref{tab_data_1}, where each parameter $\theta_i$, when varied individually with the other two set to zero, is bounded by the requirement that the standard fibre-inflation plateau not be significantly distorted. When all three
\(\theta_i\) contributions are included simultaneously, correlations
and partial cancellations can allow individual coefficients outside
the corresponding one-parameter ranges in Table~\ref{tab_data_1}.
\begin{table}[H]
\centering
\small
\setlength{\tabcolsep}{4pt}
\begin{tabular}{c|ccc}
\toprule
$|\lambda|$
& \shortstack{ $\theta_1$}
& \shortstack{$\theta_2$}
& \shortstack{ $\theta_3$}
\\
\midrule
$0$
& $[-10^{-6},\,10^{-5}]$
& $[-10^{-5},\,10^{-5}]$
& $[-10^{-5},\,10^{-6}]$
\\
$1\times10^{-7}$
& $[-10^{-5},\,10^{-5}]$
& $[-10^{-5},\,10^{-5}]$
& $[-10^{-6},\,10^{-5}]$
\\
\bottomrule
\end{tabular}
\caption{One-parameter constraints in the BHP-complete baseline of Example~1.}
\label{tab_data_1}
\end{table}

At this benchmark point, \eqref{eq:bench1}, $|A| \simeq 0.0017$, $|B|_* \simeq 0.098$, $|B|_{\rm end} \simeq 0.033$, and we characterise the largest relative deformations realised in the full accepted multi-parameter scan of \eqref{eq:smallscan}: $\theta_i=\{0,\pm10^j\}$, with $j\in[-7,7]$. The result of this multi-parameter scan is summarised in Table~\ref{tab_data_2}. For each accepted point,
the ratios are evaluated at both horizon exit and the end
of inflation, and the largest value is retained:
\bea
|\theta_1|/|A| <   5.9 \times 10^{-3}, \quad |\theta_2|/|B| < 3 \times 10^{-3}, \quad |\theta_3|/|B| < 3 \times 10^{-4}.
\label{eq:ratio_1}
\eea
It shows that a single additional $\alpha^{\prime}$-loop contribution must be sufficiently suppressed to preserve the characteristic structure of the standard fibre-inflation potential. 
The logarithmic term is especially constrained because it contributes to the same exponential falloff as \(\theta_2\), but also modifies the slope and curvature through the logarithmic factor.
The one-dimensional intervals in Table~\ref{tab_data_1}
and the deformation ratio in \eqref{eq:ratio_1} therefore summarise two distinct
analyses: the former quantify isolated-operator sensitivity, whereas
the latter characterise the largest correlated deformation realised in
the local multi-parameter scan.
\begin{table}[H]
\centering
\small
\setlength{\tabcolsep}{2pt}
\begin{tabular}{c|cccccccccc}
\toprule
Scenario & $|\lambda|$ & $\theta_1$ & $\theta_2$ & $\theta_3$ 
& $\langle \tau_f\rangle$ & $N_e$ & $n_s$ & $r$ & $\delta_{\mathcal{V}}$ & $V_{\rm corr}/V_{\rm LVS}$ \\
\midrule
BHP+$\alpha^\prime$-loop 
& $0$ 
& $10^{-6}$ & $-10^{-4}$ & $10^{-5}$ 
& $50.68$ & $60.47$ & $0.9616$ & $2.7 \times 10^{-5}$ & 0.050 & 0.06 \\
BHP+$\alpha^\prime$-loop+$F^4$ 
& $1\times 10^{-7}$ 
& $10^{-5}$ & $10^{-4}$ & $-10^{-5}$ 
& $51.66$ & $62.45$ & $0.9674$ & $2.6 \times 10^{-5}$ & 0.049 &0.06 \\
\bottomrule
\end{tabular}
\caption{Benchmark points of multi-parameter scans in the BHP-complete baseline of Example~1.}
\label{tab_data_2}
\end{table}

The inflationary energy scale
is 
\bea
V_{\rm inf}^{1/4} \sim (8 \times 10^{-13})^{1/4}M_p \simeq 2.3\times10^{15}\,\mathrm{GeV}.
\eea
In addition, we can verify that the value of the correction-sector potential at the deformed minimum remains close to that of the original BHP reference vacuum. Denoting the undeformed correction potential by $V_{\rm corr,0}$, we have
\begin{equation}
    \left|
    \frac{
        V_{\rm corr}(\langle\tau_f\rangle;\theta_1,\theta_2,\theta_3)
        -
        V_{\rm corr,0}(\langle\tau_{f,0}\rangle)
    }{
        V_{\rm corr,0}(\langle\tau_{f,0}\rangle)
    }
    \right|
    <0.05 .
\end{equation}
This result shows that the $\alpha^\prime$-loop sector is indeed   a controlled deformation of the original BHP-stabilised fibre vacuum, rather than a correction which moves the system to a qualitatively different minimum. 

As a consistency check of the sequential stabilisation procedure, we
also minimise the combined potential $V_{\rm LVS}+V_{\rm corr}$ with
respect to \(\{\tau_s,\tau_f,\mathcal V\}\). The resulting moduli
displacements are
\bea
|\lambda|=0: & \frac{|\delta \tau_s|}{\tau_s} \simeq 0.0042, \quad  & \frac{ |\delta \mathcal{V}|}{\mathcal{V}} \simeq 0.0386, \nonumber\\
|\lambda| = 10^{-7}: & \frac{|\delta \tau_s|}{\tau_s} \simeq 0.004, \quad  & \frac{|\delta \mathcal{V}|}{\mathcal{V}} \simeq 0.0367.
\eea 
which satisfy \eqref{eq:heavy-field-shift} and support the single-field treatment used above.

\subsubsection*{ KK-only BHP baseline}

In the second D7/O7 configuration of the same Calabi--Yau, the winding-type BHP correction is absent. The only conventional BHP contribution is a KK-type term of the form
\begin{equation}
    V_{\rm BHP}^{\rm KK} = \kappa \frac{W_0^2}{\mathcal{V}^2} \frac{C\tau_f}{\mathcal{V}^2},\qquad
    C = \frac{(g_s C_b^{\rm KK})^2}{72}.
\end{equation}
The resulting potential is obtained from \eqref{inf_1} by setting \(A=B=0\) and using a different value of $C$:
\begin{align} \label{inf_3}
    V_{\text{inf}} =& \kappa \frac{W_0^2}{\mathcal{V}^2} \bigg\{\mathcal{C}_{\text{dS}} + \frac{\theta_1}{\langle\tau_f\rangle^2}e^{-2k\hat{\phi}} - \frac{\theta_3 (\ln\langle\tau_f\rangle + k\hat{\phi} - 3/2)}{\mathcal{V}\sqrt{\langle\tau_f\rangle}}e^{-k\hat{\phi}/2} - \frac{ \theta_2}{\mathcal{V} \sqrt{\langle\tau_f\rangle}}e^{-k\hat{\phi}/2} \nonumber \\
    &+ \frac{2F}{3\mathcal{V}\left<\tau_f\right>}e^{-k\hat{\phi}} + \frac{{C} \langle\tau_f\rangle}{\mathcal{V}^2}e^{k\hat{\phi}} + \frac{F \sqrt{\langle\tau_f\rangle}}{\mathcal{V}^2}e^{k\hat{\phi}/2} \bigg\}\,.
\end{align}

We perform a systematic multi-parameter scan of \eqref{eq:parameters} over the range \eqref{eq:parameter-range}, where $C_{\rm BHP}^{(a)} \equiv C_b^{\rm KK}$ in the KK-only case.
When  the $\alpha'$-loop parameters are set to zero $\theta_i = 0$,
within the scanned parameter range, no viable slow-roll
solution is found when the fibre potential contains both the KK-type
 contribution and the $F^4$ correction, i.e., $V_{\rm corr} = V_{\rm BHP}^{\rm KK} + V_{F^4}$. This branch either fails to generate a
sufficient number of e-folds or gives phenomenologically unacceptable
values of $(n_s,r)$. The reason is that the required hierarchy between
the growing and decreasing exponentials is too restrictive when the
relevant terms are controlled by the same $F^4$ coefficient
$|\lambda|$.

By contrast, switching on the $\alpha^\prime$-loop sector provides the
additional decreasing structures required to generate a viable
inflationary plateau, both with and without the $F^4$ correction. We
choose the following  benchmark parameters with $a_s = 2\pi$:
\begin{align}
 C_b^{\rm KK} &= 1.415 ,\quad W_0 = 74.87 ,\quad A_s = 3.936 ,\quad  g_s = 0.4962, \nonumber \\
  \theta_1 &= 1.0\times10^{-5} ,\quad \theta_2 = 1.0\times10^{-2} ,\quad \theta_3 = -1.0\times10^{-2}.
  \label{eq:kk-only-benchmark}
\end{align}
The leading LVS minimum gives
\begin{align}  
  \langle\tau_s\rangle = 1.31 ,\quad 
 \mathcal V = 4308.84 .
\end{align}

At this benchmark point, we realise the viable $\alpha'$-loop-assisted fibre-inflation scenarios shown in Table~\ref{tab:kk-only-benchmark}. For the $|\lambda|=0$ and $|\lambda|=10^{-7}$ branches, the
corresponding field values are
\begin{equation}
 (\hat\phi_{\rm end},\hat\phi_*)
 \simeq(0.8467,3.39)
 \qquad\text{and}\qquad
 (0.8463,3.35),
\end{equation}
respectively. For the two branches, the inflationary energy scale is
$V_{\rm inf}^{1/4} \sim (7.5 \times 10^{-12})^{1/4}M_p \simeq 4\times10^{15}\,\mathrm{GeV}$. In both cases, $\epsilon_{\alpha^\prime}\simeq1.5\times10^{-4}$,
confirming that the large-volume expansion remains under control.
\begin{table}[H]
\centering
\small
\setlength{\tabcolsep}{3pt}
\begin{tabular}{c|cccccccccc}
\toprule
Scenario & $|\lambda|$ & $\theta_1$ & $\theta_2$ & $\theta_3$
& $\langle\tau_f\rangle$ & $N_e$ & $n_s$ & $r$
& $\delta_{\mathcal V}$ & $V_{\rm corr}/V_{\rm LVS}$ \\
\midrule
KK+$\alpha^\prime$-loop
& $0$
& $10^{-5}$ & $10^{-2}$ & $-10^{-2}$
& $2.94$ & $53.8$ & $0.9681$
& $2.49\times10^{-4}$ & $0.043$ & $0.055$ \\
KK+$\alpha^\prime$-loop+$F^4$
& $1\times10^{-7}$
& $10^{-5}$ & $10^{-2}$ & $-10^{-2}$
& $3.00$ & $51.7$ & $0.9622$
& $2.38\times10^{-4}$ & $0.042$ & $0.050$ \\
\bottomrule
\end{tabular}
\caption{Benchmark points of multi-parameter scans in the KK-only baseline of Example~1.}
\label{tab:kk-only-benchmark}
\end{table}

Table~\ref{tab:kk-only-benchmark} shows that viable slow-roll inflation
can be realised both with and without the $F^4$ correction. Since the
winding-type BHP contribution is absent in this configuration, the
$\theta_2$ and $\theta_3$ terms provide the essential decreasing
contribution required to form the inflationary plateau. The
$\alpha^\prime$-loop sector therefore acts as a leading ingredient of
the KK-only solution rather than as a small deformation of a complete BHP potential; thus, the deformation ratio in \eqref{eq:ratio_1} does not apply.

An a posteriori minimisation of the full potential yields
the moduli-control condition \eqref{eq:heavy-field-shift}:
\bea
\lambda=0: & \frac{|\delta \tau_s|}{\tau_s} \simeq 0.0018, \quad  & \frac{ |\delta \mathcal{V}|}{\mathcal{V}} \simeq 0.0157, \nonumber\\
|\lambda| = 10^{-7}: & \frac{|\delta \tau_s|}{\tau_s} \simeq 0.0017, \quad  & \frac{|\delta \mathcal{V}|}{\mathcal{V}} \simeq 0.0147.
\eea 
The heavy-field backreaction is therefore small in both branches, supporting the sequential stabilisation
approximation.

\subsubsection{\texorpdfstring{Example 2:  Calabi--Yau threefold with $h^{1,1}=4$}{Example 2: Calabi--Yau threefold with h11 = 4}}
\label{sec:example2-winding-alpha-loop-f4}

Consider a K3-fibred Calabi--Yau orientifold with \((h^{2,1},h^{1,1})=(98,4)\) and \(\chi=-188\) ~\cite{Cicoli:2017axo}, 
\begin{table}[H]
  \centering
 \begin{tabular}{|c||cccccccc|}
\hline
     & $x_1$  & $x_2$  & $x_3$  & $x_4$  & $x_5$ & $x_6$  & $x_7$ & $x_8$   \\
    \hline
 4 & 0 & 0  & 0 & 1 & 1 & 0 & 0  & 2   \\
 4 & 0 & 0  & 1 & 0 & 0 & 1 & 0  & 2   \\
 4 & 0 & 1  & 0 & 0 & 0 & 0 & 1  & 2   \\   
 8 & 1 & 0  & 0 & 1 & 0 & 1 & 1  & 4   \\   
 \hline
 & dP$_7$ & NdP$_{11}$  &  NdP$_{11}$ & K3 & NdP$_{11}$ & K3 & K3 & SD \\
 \hline
 \end{tabular}
 \end{table}
 with the SR ideal:
\[
{\rm SR} =  \{ x_1 x_4, \, x_1 x_6,\, x_1 x_7, \, x_2 x_7, \, x_3 x_6, \, x_4 x_5 x_8, \, x_2 x_3 x_5 x_8 \}\,. 
\]
The individual divisors are del Pezzo surfaces $dP_{7,11,11,11}$ for $D_{1,2,3,5}$, three K3 surfaces for $D_{4,6,7}$ and one special deformation divisor $D_8$. The simplified overall Calabi--Yau volume can be expressed as
\begin{equation} \label{Volume_1}
    \mathcal{V}=\frac{1}{\sqrt{2\alpha}}\sqrt{\tau_{7}}\tau_{6}-\frac{1}{3}\tau_{1}^{3/2}.
\end{equation}
where the two-cycle volume relation $t_4 \equiv \alpha\, t_6$ has been used \cite{Cicoli:2017axo}. The fibre modulus is denoted by $\tau_f\equiv\tau_7$, while the blow-up modulus $\tau_1$ is denoted by $\tau_s$; the former defines the light direction suitable for inflation.

It was shown in \cite{Altman:2021pyc,Cao:2024oqx} that there is only one non-trivial identical-divisor (NID) exchange involution, $\sigma: \{x_2 \leftrightarrow x_3, x_6 \leftrightarrow x_7 \}$, which exchanges K3 and $dP_{11}$ divisors simultaneously, while various reflection (REF) involutions can also be performed.
Here we consider two brane configurations under the same reflection involution $\sigma: \{x_8 \leftrightarrow -x_8\}$, for which there is no O3-plane but a single O7-plane on $D_8$:
\begin{itemize}
    \item Winding-only BHP case \cite{Cicoli:2017axo}: the D7-branes are not placed entirely on top of the O7-plane, with $8[O7] \equiv 8([D_8]) = 16([D_2]+[D_4]+[D_6])$, and D3-tadpole cancellation gives $|Q_{D3}| = 38$. In this case, winding-type corrections arise from brane--brane and brane--O7 intersections, while KK-type corrections are absent.
    \item BHP-absent case: when all eight D7-branes coincide with the O7-plane, the BHP corrections vanish. In this case, only $\alpha^{\prime}$-loop corrections and higher-derivative $F^4$ terms contribute to the scalar potential, as discussed in the next section.
\end{itemize}

\subsubsection*{ Winding-only BHP baseline}

In this configuration, the BHP corrections contain only the winding-type contributions:
\begin{equation} \label{V_gs_w}
    V_{W} = -\kappa\frac{|W_0|^2}{\mathcal{V}^3}\left[\frac{1}{\sqrt{\tau_f}}\left(C_W-\tilde{C}_W(\tau_f)\right)-\frac{\tau_f}{\mathcal{V}}\left(|C_3^W|-\hat{C}_W(\tau_f)\right)\right].
\end{equation}
In addition, the higher-derivative $F^4$ correction~\eqref{eq:F4-correction} to the potential is
\begin{equation} \label{eq:f4-example2}
V_{F^4}=-24\kappa^2\frac{\lambda|W_0|^4}{g_s^{3/2}\mathcal{V}^3}\left(\frac{(\alpha+1)}{\sqrt{2\alpha}}\frac{\sqrt{\tau_f}}{\mathcal{V}}+\frac{1}{\tau_f}\right),
\end{equation}

The total scalar potential in \eqref{eq:inf-potential} simplifies to
$V_{\text{corr}} = V^W_{BHP} + V_{F^4} + V_{\rm \alpha'-loop}$,
and, after canonical normalisation with $k=\frac{2}{\sqrt{3}}$, the scalar potential for the inflaton becomes
 \begin{align} \label{inf_2}
    V_{\text{inf}} &=  \kappa \frac{|W_0|^2}{\mathcal{V}^2}\bigg\{ \mathcal{C}_{\text{dS}} + \frac{\theta_1}{\langle\tau_f\rangle^2}e^{-2k\hat{\phi}}- \frac{B + \theta_2}{\mathcal{V}\sqrt{\langle\tau_f\rangle}}e^{-\frac{k}{2}\hat{\phi}} - \frac{\theta_3 (\ln\langle\tau_f\rangle + k\hat{\phi} - 3/2)}{\mathcal{V}\sqrt{\langle\tau_f\rangle}}e^{-\frac{k}{2}\hat{\phi}} + \frac{C\langle\tau_f\rangle}{\mathcal{V}^2}e^{k\hat{\phi}}  \nonumber \\
    &+ \frac{F_1}{\mathcal{V}\langle\tau_f\rangle}e^{-k\hat{\phi}} + \frac{F_2\sqrt{\langle\tau_f\rangle}}{\mathcal{V}^2}e^{\frac{k}{2}\hat{\phi}}   \bigg\},
\end{align}
where the parameters are defined as in \cite{Cicoli:2017axo}:
\bea
 F_1 = \frac{3}{\pi}\frac{|\lambda|W_0^2}{\sqrt{g_s}} ,\quad B = C_W - \tilde{C}_W ,\quad F_2 = \frac{\alpha + 1}{\sqrt{2\alpha}}  F_1 ,\quad C = |C_3^W| - \hat{C}_{W},
 \label{eq:winding-only-parameter}
\eea
with coefficients
\bea
& C_W = \sqrt{2\alpha} \left(C_1^W + \frac{C_2^W}{\alpha}\right) , \qquad \tilde{C}_W = \frac{ |C_4^W|}{(1 + \alpha)}\sqrt{\frac{\alpha}{2}}\left(1 - \frac{\sqrt{2\alpha}}{\alpha + 1}\sqrt{\frac{\langle\tau_s\rangle}{\tau_f}}\right)^{-1}, \nonumber\\
& \hat{C}_W = \frac{C_5^W}{2}\left(1 + \frac{1}{\sqrt{2\alpha}}\frac{\tau_f^{3/2}}{\mathcal{V}}\right)^{-1} + \frac{C_{6}^W}{2}\left(1 + \sqrt{\frac{\alpha}{2}}\frac{\tau_f^{3/2}}{\mathcal{V}}\right)^{-1}. 
\eea
Similarly, $C_1^{\rm W}$, $C_2^{\rm W}$, $C_4^{\rm W}$, $C_5^{\rm W}$ and $C_6^{\rm W}$ parametrize the winding-loop contributions associated with the relevant D7/O7 intersection curves involving the divisors $D_2$, $D_4$, $D_6$ and $D_8$.
For $\alpha=1$, the K\"ahler-cone condition along the effective fibre
direction derived in~\cite{Cicoli:2017axo} is
\begin{equation}
     \frac{\sqrt{3}}{2}\ln\left(\frac{2\langle\tau_s\rangle}{\langle\tau_f\rangle}\right)< \hat{\phi}<\frac{\sqrt{3}}{2}\ln\left(\frac{\mathcal{V}}{\langle\tau_f\rangle\sqrt{\langle\tau_s\rangle}}\right)\,\
     \label{eq:kahlercone2}
\end{equation}

We choose the following benchmark point to achieve consistent winding-driven inflation with $a_s = 2\pi$\footnote{We do not use the benchmark values of \cite{Cicoli:2017axo}, since they do not satisfy the latest observational constraints.}:
 \begin{align}\label{set}
	& C_1^W = 0.33 ,\, C_2^W = 0 ,\, |C_3^W| = 7 \times 10^{-5} ,\, |C_4^W| = 1.07 ,\, C_5^W = -0.1287 \nonumber \\ &C_6^W = 0 ,\, 
	A_s = 13 ,\, W_0 = 27 ,\, g_s = 0.49 .
 \end{align}
The leading LVS minimum gives
\begin{align}
    \left<\tau_s\right>  \simeq 1.638 \,, \quad \mathcal{V} \simeq 2892 \,.
\end{align}
At this benchmark point, we realise the winding-type BHP fibre-inflation scenarios shown in Table~\ref{tab:winding-baseline}, with heavy-modulus control parameter $\delta_\cV \simeq 0.036$, inflationary energy scale
$V_{\rm inf}^{1/4} \sim (2.32 \times 10^{-12})^{1/4}M_p \simeq 3\times10^{15}\,\mathrm{GeV}$ and expansion-control parameter $\epsilon_{\alpha'} \simeq 2.3 \times 10^{-4}$.
\begin{table}[ht]
\centering
\small
\setlength{\tabcolsep}{4pt}
\begin{tabular}{lcccccccc}
\toprule
Scenario
& $|\lambda|$
& $\langle\tau_{f,0}\rangle$
& $V_{\rm corr}/V_{\rm LVS}$
& $\hat\phi_{\rm end}$
& $\hat\phi_*$
& $n_s$
& $r$
& $N_e$
\\
\midrule
Winding-only
& $0$
& $61.94$
& $0.0777$
& $0.975$
& $3.06$
& $0.9622$
& $7.457\times10^{-5}$
& $61.0$
\\
Winding-only+$F^4$
& $1\times10^{-7}$
& $61.96$
& $0.0776$
& $0.975$
& $3.08$
& $0.9681$
& $7.456\times10^{-5}$
& $64.3$
\\
\bottomrule
\end{tabular}
\caption{Winding-only baseline in Example~2.}
\label{tab:winding-baseline}
\end{table}

Now, we turn on the  $\alpha^\prime$-loop corrections, characterized by the coefficients $\theta_i$,  and examine their effects on the winding-only  fibre potential.  
The resulting one-parameter scans are shown in
Table~\ref{tab_data_3}. The logarithmic coefficient $\theta_3$ is
typically restricted one order of magnitude more strongly than
$\theta_2$ and the inflationary energy scale is
$V_{\rm inf}^{1/4} \sim (2.3 \times 10^{-12})^{1/4}M_p \simeq 3\times10^{15}\,\mathrm{GeV}$. We then allow all three coefficients to vary simultaneously as in \eqref{eq:smallscan}.
The result of this multi-parameter scan is summarised in Table~\ref{tab_data_4}. For each accepted point,
the  largest relative deformations found in the accepted
multi-parameter sample are
\begin{align} 
	|\theta_2|/|B| \lesssim 1.7 \times 10^{-3} , \quad |\theta_3|/|B| \lesssim 1.7 \times 10^{-4}.
\end{align}
Here the values are obtained from the full accepted sample rather than
from the representative points in Table~\ref{tab_data_4}. The resulting fractional
shifts are
\bea
\frac{|\delta\tau_s|}{\tau_s}\simeq0.0006,
\qquad
\frac{|\delta\mathcal V|}{\mathcal V}\simeq0.0069
\eea
for both \(|\lambda|=0\) and 
\(|\lambda|=10^{-7}\). The heavy-field backreaction therefore remains
negligible, consistently with \eqref{eq:heavy-field-shift}.
\begin{table}[ht]
\centering
\small
\setlength{\tabcolsep}{5pt}
\begin{tabular}{cccc}
\toprule
$|\lambda|$
& \shortstack{ $\theta_1$}
& \shortstack{ $\theta_2$}
& \shortstack{ $\theta_3$}
\\
\midrule
$0$
& $[-10^{-4},\,10^{-4}]$
& $[-10^{-5},\,10^{-4}]$
& $[-10^{-6},\,10^{-5}]$
\\
$1\times10^{-7}$
& $[-10^{-4},\,10^{-4}]$
& $[-10^{-4},\,10^{-5}]$
& $[-10^{-5},\,10^{-6}]$
\\
\bottomrule
\end{tabular}
\caption{One-parameter constraints in the winding-only baseline of Example~2.}
\label{tab_data_3}
\end{table}

\begin{table}[!t]
\small 
\centering
\setlength{\tabcolsep}{2pt}
\begin{tabular}{cccccccccc}
\toprule
Scenario & $|\lambda|$ & $\theta_1$ & $\theta_2$ & $\theta_3$ 
& $\langle \tau_f\rangle$ & $N_e$ & $n_s$ & $r$ & $\delta_{\mathcal V}$ \\
\midrule
Winding+$\alpha^\prime$-loop 
& $0$ 
& $10^{-4}$ & $10^{-4}$ & $-10^{-5}$ 
& 63.7 & 62.7 & 0.9653 & $7.3 \times 10^{-5}$ & 0.036 \\
Winding+$\alpha^\prime$-loop+$F^4$ 
& $1 \times 10^{-7}$ 
& $10^{-4}$ & $-10^{-4}$ & $-10^{-5}$ 
& 63.8 & 60.4 & 0.9611 & $7.3 \times 10^{-5}$ & 0.036  \\
\bottomrule
\end{tabular}
\caption{Benchmark points of multi-parameter scans in the winding-only baseline of Example~2.}
\label{tab_data_4}
\end{table}

\subsubsection{Summary of the BHP-driven constraints}

We summarise the constraints on the generalised
\(\alpha'\)-loop sector for different BHP baselines in
Table~\ref{tab:conclusion-BHP-baselines}. The main lesson is that the
meaning of the \(\theta_i\) coefficients depends on the underlying BHP
structure. When the corresponding BHP operators are present, the
\(\alpha'\)-loop sector acts as a controlled deformation of the fibre
potential. The accepted scans of Examples~1 and~2 give
\bea
\mathcal R_{\alpha^\prime/{\rm BHP}}
\lesssim \mathcal O(10^{-2}),
\eea
where $R_{\alpha'/\mathrm{BHP}}$ denotes the largest coefficient ratio
relative to a matched nonzero BHP coefficient, evaluated at horizon exit
and at the end of inflation. The logarithmic coefficient is more tightly
bounded, with
\bea
|\theta_3/B|\lesssim \mathcal O(10^{-3}).
\eea
These bounds should be understood as benchmark-dependent deformation
bounds rather than universal microscopic constraints. For the two mixed $\alpha'$-loop terms,
$|V_{\log}/V_{\theta_2}|=
|(\theta_3/\theta_2)(\ln\tau_f-3/2)|$.

When a BHP channel is absent, the corresponding \(\alpha'\)-loop
structure no longer represents a fractional deformation of an existing
operator. In the KK-only configuration, the missing decreasing
contributions are supplied by the \(\theta_2\) and \(\theta_3\) terms,
allowing viable inflation where the KK contribution alone supplemented
by \(F^4\) fails. Indeed, the KK-only benchmarks in Table~4 have
$|V_{\log}/V_{\theta_2}|_*\simeq 3.50$ and $3.47$, respectively,
so the logarithmic contribution exceeds the non-logarithmic mixed term
at horizon exit. In the winding-only configuration, the \(\theta_1\)
term provides the missing KK-like structure, while \(\theta_2\) and
\(\theta_3\) deform the existing winding contribution. Thus the same
\(\alpha'\)-loop sector can interpolate between deformation-dominated
and correction-assisted inflation depending on the BHP baseline.

Finally, the logarithmic coefficient is particularly constrained by the
flatness requirements of an existing BHP plateau. In the simultaneous
scans, its log-enhanced contribution can nevertheless be comparable to
or larger than the non-logarithmic $\theta_2$ term at horizon exit.
Correlations and partial cancellations can also allow individual
coefficients outside their single-coefficient intervals. In the fully
BHP-free regime studied in Section~\ref{sec:alpha-loop-inflation}, the logarithmic coefficient instead
becomes an independent shape parameter of the leading inflaton potential.

\begin{table}[ht]
\centering
\small
\renewcommand{\arraystretch}{1.25}
\begin{tabular}{lcccc}
\toprule
BHP baseline
& Conventional structure
& $|\theta_1/A|$
& $|\theta_2/B|$
& $|\theta_3/B|$
\\
\midrule
BHP-complete
& \(A,B,C\neq0\)
& $<\cO(10^{-2})$
& $<\cO(10^{-2})$
& $<\cO(10^{-3})$
\\[1mm]

KK-only
& \(A=B=0,\ C\neq0\)
& -- 
& --
& \text{Not applicable}
\\[1mm]

Winding-only
& \(A=0,\ B, C\neq0\)
& --
& $<\cO(10^{-2})$
& $<\cO(10^{-3})$
\\
\bottomrule
\end{tabular}
\caption{
Benchmark-dependent upper bounds on fractional
\(\alpha'\)-loop deformations. A dash denotes an undefined ratio.
}
\label{tab:conclusion-BHP-baselines}
\end{table}

\section{\texorpdfstring{$\alpha'$-loop inflation in the BHP-free regime}{alpha-prime-loop inflation in the BHP-free regime}}
\label{sec:alpha-loop-inflation}

A BHP-free configuration can arise when the D7/O7 arrangement removes
both the conventional KK- and winding-type BHP channels. 
This provides a qualitatively different realization from standard
fibre inflation, where the plateau is generated by the BHP loop
potential.
The remaining
fibre potential is then generated by the generalized $\alpha'$-loop
sector, possibly supplemented by the higher-derivative $F^4$
correction. The absence of the conventional BHP terms raises a non-trivial
question: without the usual KK- and winding-type hierarchy, what
relative sizes of the remaining fibre-dependent corrections are required
to stabilise the fibre modulus and generate a slow-roll plateau? The answer is encoded in a characteristic hierarchy between the
different \(\alpha'\)-loop structures, which we derive analytically
below.

\subsection{Analytic constraints}

Following the LVS hierarchy, we first stabilise the moduli
\((\mathcal V,\tau_s)\) at leading order. The remaining fibre
direction is then lifted by the subleading correction sector.
\begin{equation}
 V_{\rm corr}
 =
 \kappa\frac{|W_0|^2}{\mathcal V^2}
 \left[
 \frac{\theta_1}{\tau_f^2}
 -\frac{\theta_2}{\mathcal V\sqrt{\tau_f}}
 -\frac{\theta_3(\ln\tau_f-3/2)}
        {\mathcal V\sqrt{\tau_f}}
 +\frac{F_1}{\mathcal V\tau_f}
 +\frac{F_2\sqrt{\tau_f}}{\mathcal V^2}
 \right].
 \label{eq:bhp-free-full-potential}
\end{equation}
For the analytic determination of the fibre minimum we neglect the
$F^4$ terms at $\hat\phi=0$, since the coefficient
of the growing exponential is strongly suppressed there. This approximation
is justified by its small relative contribution to the fibre
stationarity equation at the minimum, while the full \(F^4\) potential
is retained in the numerical inflationary evolution. Setting $F_1=F_2=0$ then gives the leading potential
\begin{equation}
 V_{\rm BHP-free}
 =
 \kappa\frac{|W_0|^2}{\mathcal V^2}
 \left[
 \mathcal C_{\rm dS}
 +\frac{\theta_1}{\tau_f^2}
 -\frac{\theta_2}{\mathcal V\sqrt{\tau_f}}
 -\frac{\theta_3(\ln\tau_f-3/2)}
        {\mathcal V\sqrt{\tau_f}}
 \right],
 \label{eq:bhp-free-leading-potential}
\end{equation}
where the constant $\mathcal C_{\rm dS}$ is chosen such that
$V_{\rm BHP-free}(\langle\tau_f\rangle)=0$.
The stationary condition \(\partial_{\tau_f}V_{\rm BHP-free}=0\)  gives
\bea
 \frac{\theta_1}{\theta_2}
 =
 \frac{\langle\tau_f\rangle^{3/2}}{4\mathcal V}
 \left[1+ \frac{\theta_3}{\theta_2}(\ln\langle \tau_f \rangle-\frac{7}{2})\right], \qquad
 \frac{\mathcal C_{\rm dS}}{\theta_2}
 \mathcal V\sqrt{\langle\tau_f\rangle}
 =
 \frac34\left[1+\frac{\theta_3}{\theta_2}\left(\ln\langle \tau_f \rangle -\frac56\right)\right].
 \label{eq:stationary-free}
\eea
This first relation is the characteristic hierarchy of the BHP-free regime.
It
shows that the relative size of the fibre-scale and mixed
\(\alpha'\)-loop structures is not arbitrary, but is fixed by the LVS
volume hierarchy up to the logarithmic deformation.
For convenience, we define the parameter ratios
\begin{equation}
 r_1 \equiv\frac{\theta_1}{\theta_2},
 \qquad
 r_3 \equiv \frac{\theta_3}{\theta_2},
 \qquad
 L_0 \equiv \ln\langle\tau_f\rangle-\frac32,
\end{equation}
The local-minimum condition then gives
$
 \theta_2\left[3+r_3(3L_0-4)\right]>0.
$
For the numerical branch studied below, $\theta_2>0$ and
$\langle\tau_f\rangle<e^{17/6} \simeq 17$, so $3L_0-4<0$ and we obtain the local-minimum constraint
\bea
\label{eq:local-minimum-ratio}
 r_3<\frac{3}{4-3L_0}.
\eea
This condition constrains the logarithmic deformation independently of
the overall potential scale, since an excessively positive \(r_3\) changes the
curvature of the fibre potential and removes the local minimum.

Using
$\tau_f=\langle\tau_f\rangle e^{k\hat\phi}$ with
$k=2/\sqrt3$, the inflationary potential becomes
 \begin{align} \label{inf_4}
    V_{\text{inf}} =& \kappa \frac{|W_0|^2}{\mathcal{V}^2}\bigg\{ \mathcal{C}_{\text{dS}} +  \frac{\theta_1}{\langle\tau_f\rangle^2}e^{-2k\hat{\phi}}- \frac{\theta_2}{\mathcal{V}\sqrt{\langle\tau_f\rangle}}e^{-\frac{k}{2}\hat{\phi}} - \frac{\theta_3 (L_0 + k\hat{\phi})}{\mathcal{V}\sqrt{\langle\tau_f\rangle}}e^{-\frac{k}{2}\hat{\phi}} \bigg\} \,.
\end{align}
Define
\begin{align}
D(\hat \phi) \equiv & \,\,3L_1 e^{k \hat \phi/2} + L_2 e^{-3k \hat\phi/2} - 4 - 4r_3(L_0 + k \hat \phi) ,\nonumber \\
S_* \equiv & \,\, \frac{3}{4}L_1 + \frac{1}{4}L_2 e^{-2k\hat \phi_*} - [1+r_3(L_0 + k\hat \phi_*)]e^{-k \hat \phi_*/2} \,,
\end{align}
with $L_1 = 1 + r_3(L_0 + 2/3)$ and $L_2 = 1 + r_3(L_0 -2)$. Using \eqref{eq:stationary-free} to express $r_1$ in terms of $r_3$, the exact potential slow-roll parameters of the truncated \(\lambda = 0\) model
\bea
\epsilon_V(r_3, \hat \phi) =& \frac{\bigg[-2k L_2 e^{-3k\hat{\phi}/2} + 2k + 2k r_3 (L_0 - 2 + k\hat{\phi}) \bigg]^2}{2D(\hat \phi)^2}, \,\nonumber\\
\eta_V(r_3, \hat \phi) =& \frac{4k^2 L_2 e^{-\frac{3k}{2}\hat{\phi}} - k^2 - k^2 r_3(L_0 - 4 + k\hat{\phi}) }{ D(\hat \phi)}.
\eea
Therefore, the cosmological constraints \eqref{power_spectrum}--\eqref{eq:cosmological}, heavy-modulus control \eqref{eq:heavy-modulus-hierarchy} and the K\"ahler-cone condition \eqref{eq:kahlercone2} can be rewritten in terms of the LVS parameters and $r_3$. For example, the scalar-amplitude condition \eqref{power_spectrum} determines $\theta_2$ at $\hat\phi_*$ through
\begin{align}
 \theta_2
 =
 \frac{
  2.1\times10^{-9}\,
  24\pi^2\epsilon_{V}(r_3, \hat \phi_*)\,
  \mathcal V^3\sqrt{\langle\tau_f\rangle}
 }{
  \kappa|W_0|^2S_*(r_3, \hat \phi_*)
 }.
 \label{eq:horizon-condition}
\end{align}
Substituting \eqref{eq:horizon-condition} into the heavy-modulus condition
\eqref{eq:heavy-modulus-hierarchy} gives an additional constraint on the LVS  parameters and $r_3$,
\begin{equation}
 \frac{
  4(2.1\times10^{-9})\,24\pi^2
  \epsilon_{V}(r_3, \hat \phi_*)\mathcal V^3
 }{
  9\kappa|W_0|^2\hat\xi
 }
 <10^{-2}.
 \label{eq:bhp-free-heavy-control}
\end{equation}

These constraints provide an analytic procedure for fixing the parameters within a narrow window.
For fixed
$\mathcal V$ and trial $(r_1,r_3)$, all positive roots for $\langle\tau_f\rangle$ in \eqref{eq:stationary-free} are determined. These roots are then
tested against \eqref{eq:local-minimum-ratio} and the K\"ahler-cone bounds \eqref{eq:kahlercone2}.
The slow-roll conditions constrain the shape parameters, while
\eqref{eq:horizon-condition} determines $\theta_2$. The individual
coefficients are then reconstructed as
$\theta_1=r_1\theta_2$ and $\theta_3=r_3\theta_2$.

\subsection{Numerical windows of parameters}

We scan the parameters in the range~\eqref{eq:parameter-range} with $C_{\rm BHP}^{(a)}=0$, and solve the LVS conditions for
\((\tau_s,\mathcal V)\) and subsequently determine the fibre minimum and the
inflationary trajectory. The cosmological constraints \eqref{power_spectrum}--\eqref{eq:cosmological}, large-volume expansion \eqref{eq:epsilon}--\eqref{eq:ratio}, heavy-modulus control \eqref{eq:heavy-modulus-hierarchy} and the K\"ahler-cone condition \eqref{eq:kahlercone2} jointly restrict the $\alpha'$-loop sector and the LVS parameters to the windows shown in Table~\ref{tab:bhp-free-scan-summary}.
\begin{table}[ht]
\centering
\small
\renewcommand{\arraystretch}{1.15}
\begin{tabular}{@{}lp{0.70\textwidth}@{}}
\toprule
Parameters & Constraints \\
\midrule

 $F^4$ coefficient
& $|\lambda|=0$ \\

$\alpha'$-loop coefficients
& $1.6\times10^{-6}\lesssim\theta_1\lesssim3.9\times10^{-5}$,\newline
  $4.0\times10^{-3}\lesssim\theta_2\lesssim1.8\times10^{-2}$,\newline
  $-2.2\times10^{-3}\lesssim\theta_3\lesssim-2.0\times10^{-4}$
\\

LVS parameters
& $40\lesssim W_0\lesssim500$, \quad
  $35\lesssim A_s\lesssim100$, \quad
  $0.453\lesssim g_s\lesssim0.50$ \\

\midrule

$F^4$ coefficient
& Main viable branch at $|\lambda|=10^{-7}$ \\

$\alpha'$-loop coefficients
& $1.1\times10^{-5}\lesssim\theta_1
   \lesssim2.2\times10^{-5}$,\newline
  $9.5\times10^{-3}\lesssim\theta_2
   \lesssim1.45\times10^{-2}$,\newline
  $-3.1\times10^{-3}\lesssim\theta_3
   \lesssim-1.8\times10^{-3}$ \\

LVS parameters
& $48\lesssim W_0\lesssim75$, \quad
  $39\lesssim A_s\lesssim67$, \quad
  $0.470\lesssim g_s\lesssim0.500$ \\

\bottomrule
\end{tabular}
\caption{Valid windows for parameters in $\alpha'$-loop inflation.}
\label{tab:bhp-free-scan-summary}
\end{table}

\begin{figure}[ht]
 \centering
 \includegraphics[width=0.47\textwidth]{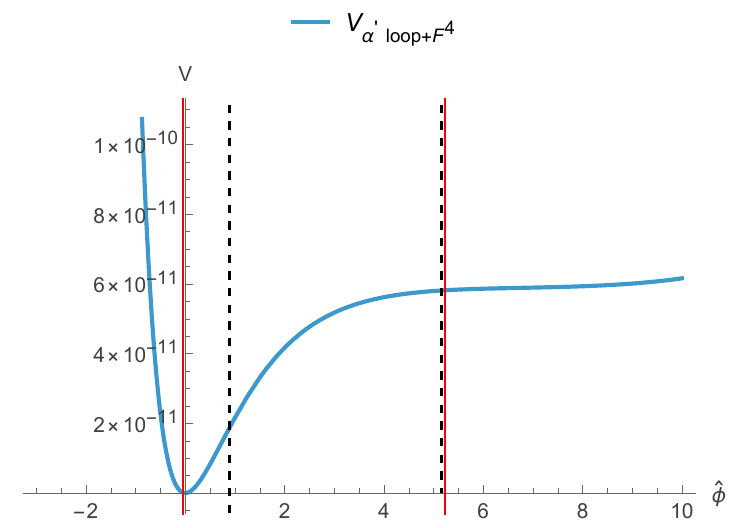}
 \hfill
 \includegraphics[width=0.47\textwidth]{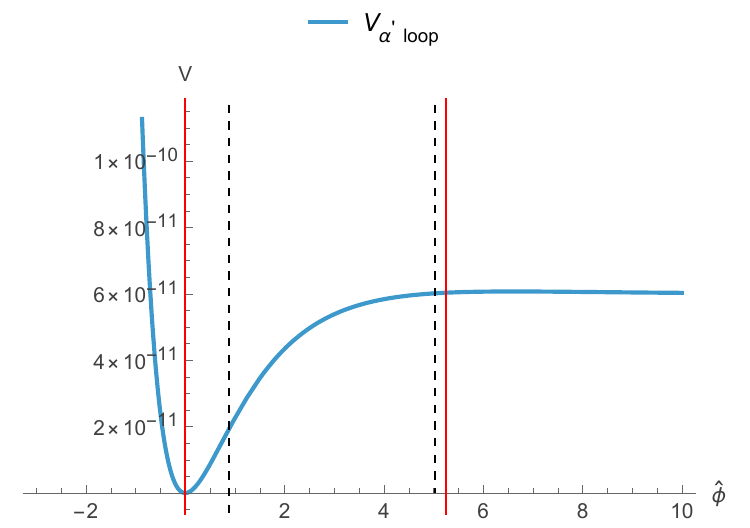}
 \caption{$\alpha'$-loop inflationary potentials for benchmark points of
 $\eqref{alpha-loop-parameter-choice}$, with and without the $F^4$ correction.}
 \label{fig:alpha-loop-f4}
\end{figure}

For illustration, consider the benchmark  parameter choice
\begin{equation}
 \begin{gathered}
 W_0=66.3,\qquad A_s=52.5,\qquad
 g_s=0.4877,\qquad a_s=2\pi,\\
 \theta_1=1.376\times10^{-5},\qquad
 \theta_2=1.163\times10^{-2},\qquad
 \theta_3=-2.18\times10^{-3}.
 \end{gathered}
 \label{alpha-loop-parameter-choice}
\end{equation}
Then the leading LVS minimum and expansion parameter are 
\bea
 \langle\tau_s\rangle \simeq 1.65,
 \qquad
 \langle\mathcal V\rangle \simeq 1847.54,
 \qquad
 \epsilon_{\alpha'}\simeq3.6\times10^{-4}.
\eea
The corresponding inflationary solutions are summarised in
Table~\ref{tab:bhp-free-benchmark}, and the inflationary potentials are shown in Figure~\ref{fig:alpha-loop-f4}, where the red solid vertical line denotes the K\"ahler-cone bound and the black dashed lines indicate horizon exit and the end of inflation.

The associated parameter ratios $(r_1,r_3)$ are
\begin{equation}
 r_1
 =\frac{\theta_1}{\theta_2}
 =1.18315\times10^{-3},
 \qquad
 r_3
 =\frac{\theta_3}{\theta_2}
 =-0.187446.
 \label{eq:benchmark-ratios}
\end{equation}
The benchmark thus realises the large-volume suppression of
$\theta_1/\theta_2$ predicted by the stationarity condition, while
$\theta_3/\theta_2$ remains an independent shape parameter.
For the two viable solutions in Table~\ref{tab:bhp-free-benchmark},
$|V_{\log}/V_{\theta_2}|_*\simeq 1.03$ and $1.07$, respectively,
showing that logarithmic and non-logarithmic mixed contributions of
comparable magnitude coexist in a viable slow-roll potential.

\begin{table}[!ht]
\centering
\small
\begin{tabular}{lccccccccc}
\toprule
Scenario
& $|\lambda|$
& $\langle\tau_f\rangle$ & $\hat\phi_{\rm end}$ & $\hat\phi_{*}$
& $N_e$ & $n_s$ & $r$ & $V_{\rm corr}/V_{\rm LVS}$
& $\delta_{\mathcal V}$\\
\midrule
Pure $\alpha'$-loop
& $0$ & $3.34$ & 0.890 & 5.02 & $59.2$ & $0.9617$ & $1.93 \times 10^{-3}$ & $0.092$ & $0.038$\\
$\alpha'$-loop+$F^4$
& $10^{-7}$ & $3.44$ & 0.891 & 5.16 & $64.6$ & $0.9685$ & $1.87 \times 10^{-3}$ & $0.088$ & $0.038$\\
\bottomrule
\end{tabular}
\caption{Benchmark points of multi-parameter scans in the BHP-free baseline.}
\label{tab:bhp-free-benchmark}
\end{table}

We can verify that the fibre sector does not significantly displace the
heavy moduli. A direct minimisation of the combined potential gives
\bea
\lambda=0: & \frac{|\delta \tau_s|}{\tau_s} \simeq 0.0015,  \quad & \frac{ |\delta \mathcal{V}|}{\mathcal{V}} \simeq 0.0157, \nonumber\\
|\lambda| = 10^{-7}: & \frac{|\delta \tau_s|}{\tau_s} \simeq 0.0014, \quad  & \frac{|\delta \mathcal{V}|}{\mathcal{V}} \simeq 0.0149.
\eea

A local window for $r_1$ and $r_3$ can then be obtained without treating
$\langle\tau_f\rangle$ as an independent input. Fixing
$\mathcal V=1847.54$, $W_0=66.3$, $g_s=0.4877$,
$\theta_2=1.163\times10^{-2}$ and $N_e=59.2$, we vary $r_3$,
determine $\langle\tau_f\rangle$ from the scalar-amplitude and
$e$-fold conditions, and subsequently determine $r_1$ through
\eqref{eq:stationary-free}. Imposing the  K\"ahler-cone condition then gives the range of $\alpha'$-loop parameters
\begin{equation}
 -0.1934\lesssim r_3\lesssim-0.1865,
 \qquad
 1.16\times10^{-3}\lesssim r_1\lesssim1.41\times10^{-3},
 \qquad
 3.30\lesssim\langle\tau_f\rangle\lesssim3.78 .
 \label{eq:benchmark-ratio-window}
\end{equation}
The numerical ratios in \eqref{eq:benchmark-ratios} lie inside this interval.
The interval is benchmark dependent, rather than a model-independent bound,
but it demonstrates that the scanned solution is consistent with the analytic
structure of the potential.

For the $F^4$ deformation,  $\lambda=10^{-7}$, the use of the same leading fibre minimum is justified
by the suppression of the $F^4$ contribution at $\hat\phi=0$. The justification is not based only on the size of the \(F^4\)
potential itself, but on its contribution to the fibre stationarity
condition. For the
benchmark, the relative corrections to the stationary condition are
\bea
 \Delta_{F_1}
 =
 \frac{2F_1}{
 \sqrt{\langle\tau_f\rangle}
 \left|
 \theta_2+\theta_3
 \left(L_0-2\right)
 \right|}
 \simeq 3.9\times10^{-2},\,
 \Delta_{F_2}
 =
 \frac{F_2\langle\tau_f\rangle}{
 \mathcal V
 \left|
 \theta_2+\theta_3
 \left(L_0-2\right)
 \right|}
 \simeq9.2\times10^{-5}.
\eea
The growing exponential is therefore negligible at the fibre minimum, while
the remaining $F^4$ term produces only a few-percent correction to the leading
stationary condition. The full $F^4$ potential is nevertheless retained along
the inflationary trajectory, leading to the second row of
Table~\ref{tab:bhp-free-benchmark}. For the two branches,
$V_{\rm inf}(\hat\phi_*)\simeq 6.01 \times10^{-11}$ and
$5.81\times10^{-11}$, respectively, corresponding to the common
inflationary energy scale $V_{\rm inf}^{1/4}\sim 6.7\times10^{15}\,{\rm GeV}$.

Combining the analytic and numerical analyses, ~\eqref{eq:stationary-free} also explains
the observed coefficient hierarchy for generic values of $\langle\tau_f\rangle \sim \cO(1-100)$. At fixed coefficient ratios, $\theta_2$ sets the overall scale of the leading loop potential. The stationarity condition suppresses the coefficient ratio $\theta_1/\theta_2$ by the large-volume factor, while inflationary
consistency constrains the shape parameter $\theta_3/\theta_2$:
\begin{equation}
\label{eq:conclusion-theta-hierarchy}
 \theta_2 \simeq \cO(10^{-2}), \quad |\frac{\theta_3}{\theta_2}|\lesssim \cO(10^{-2})\text{--}\cO(10^{-1}), \quad \frac{\theta_1}{\theta_2}
    \simeq
    \cO(1) \frac{\langle\tau_f\rangle^{3/2}}{4\mathcal V}.
\end{equation}

\section{Conclusions and outlook}
\label{sec:con}

We have revisited fibre inflation in Type IIB Large Volume
compactifications beyond the conventional BHP parametrisation of
string-loop effects. Motivated by the EFT classification of genuine
loops, local higher-derivative operators and warping effects, we
isolated the relevant fibre-dependent sector and parametrised it as the
restricted ``\(\alpha'\)-loop'' potential (\ref{eq:alpha-loop-potential}). The coefficients \(\theta_i\) encode compactification-dependent data
and are constrained phenomenologically by perturbative control,
heavy-modulus stability and inflationary flatness. In principle, they
are fixed by the compactification, but cannot be computed explicitly
without the Calabi--Yau metric. Our analysis shows that the fibre potential is
determined not only by the Calabi--Yau volume form, but also by the
orientifold involution, O-plane content and D7-brane configuration.
This dependence allows the same fibred large-volume geometry to realise
BHP-driven, partially BHP-driven or BHP-free inflationary potentials.

Our main result is the identification of an EFT-controlled BHP-free
fibre-inflation regime, which we refer to as \(\alpha'\)-loop inflation,
in an explicit K3-fibred orientifold setup. When the conventional
KK- and winding-type BHP channels are absent, the generalized
\(\alpha'\)-loop sector can nevertheless stabilise the fibre modulus and
generate a CMB-compatible slow-roll plateau. A key feature of this
regime is that the relative size of the different fibre-dependent
structures is not an arbitrary tuning of effective coefficients. The
large-volume stationarity condition enforces a characteristic hierarchy
between the fibre-scale and mixed contributions,
\begin{equation}
\frac{\theta_1}{\theta_2}
\simeq
\frac{\langle\tau_f\rangle^{3/2}}{4\mathcal V}
\left(
1+
\frac{\theta_3}{\theta_2}
\left(
\ln\langle\tau_f\rangle-\frac72
\right)
\right).
\end{equation}
Thus, the BHP-free minimum is controlled by the hierarchy between the ratios
$\theta_1/\theta_2$ and $\theta_3/\theta_2$. The former is suppressed
by the large-volume factor
$\langle\tau_f\rangle^{3/2}/\mathcal V$, while the latter controls the
logarithmic deformation of this hierarchy. Hence, $\theta_2$ fixes the
overall potential scale, whereas these two ratios determine the
position and shape of the fibre minimum.

Once the LVS minimum is imposed, \(\mathcal V\) is tied to \(W_0\), \(g_s\), \(\xi\), \(a_s\) and \(A_s\). Combining this relation with the CMB amplitude, the K\"ahler-cone condition and the hierarchy \(H_*/m_{\mathcal V}\) gives a finite control window characterised by
\begin{equation}
    \theta_2 \simeq \cO(10^{-2}), \qquad 
    |\frac{\theta_3}{\theta_2}|
    \lesssim
    \cO(10^{-2})\text{--}\cO(10^{-1}),\qquad
    \frac{\theta_1}{\theta_2}
    \simeq
    \cO(1) \frac{\langle\tau_f\rangle^{3/2}}{4\mathcal V}.
\end{equation}
Stronger hierarchy requirements push the allowed coefficients downward. Extremely weak string coupling can enlarge the formal range, but typically requires a very large volume and flux superpotential, so flux and tadpole constraints must be checked in explicit compactifications.

The three BHP baselines considered in this work lead to qualitatively
different roles for the \(\alpha'\)-loop sector. Writing the conventional
single-field BHP potential in terms of the three structures
\(A/\tau_f^2\), \(-B/(\mathcal V\sqrt{\tau_f})\) and
\(C\tau_f/\mathcal V^2\), the BHP-complete case contains all three.
On the branch continuously connected to standard fibre inflation,
\(\theta_1\) and \(\theta_2\) renormalise the corresponding power-law
coefficients, whereas \(\theta_3\) gives a new logarithmic deformation.
The upper bounds on the allowed fractional deformations are strongly
geometry dependent and give upper bounds
\(\mathcal O(10^{-2})\), with the least restrictive cases reaching the percent level. Correlated scans also admit a
distinct logarithmically assisted BHP-complete branch. 
In KK-only configurations the mixed \(\theta_2\) and \(\theta_3\)
terms can instead be leading ingredients, while in winding-only
configurations the strongest constraints apply to their deformation of
the existing winding structure and to the new
\(\theta_1/\tau_f^2\) term.  In all three cases, the decisive
conditions are the CMB observables, the K\"ahler-cone field range, the
positivity of the covariant heavy Hessian and the small backreaction of
the LVS moduli. These results are summarised in Table~\ref{tab:conclusion-BHP-baselines}.

Several directions deserve further study. First, it would be important to determine the coefficients \(\theta_i\) microscopically. This requires detailed knowledge of the Calabi-Yau metric, orientifold projection, wrapped D7/O7 divisors, open-string spectrum and relevant string amplitudes. In particular, one should decide whether candidate marginal operators such as \(R_8^4\) on D7/O7 worldvolumes or \(R_6^3\) on intersection loci are actually present in a given compactification.

Second, the analysis should be extended to a systematic scan over explicit orientifold Calabi--Yau geometries. One should classify K3-fibred orientifolds according to their BHP channels, possible \(F^4\) contribution, K\"ahler-cone range and compatibility with D3/D7 tadpole cancellation, Freed--Witten consistency and uplift data. This would clarify how common the BHP-free regime is in the landscape of global Type IIB models.

Third, the single-field treatment should be upgraded to a full multi-field analysis. Although our controlled benchmarks keep the volume and blow-up moduli sufficiently heavy, a complete treatment should follow their displacement along the inflationary trajectory, include the moduli dependence of the uplift sector and evaluate the full Hessian throughout the path. This is also important for reheating, moduli decay, dark radiation and possible isocurvature effects.

More broadly, our results suggest that perturbative corrections beyond
the conventional BHP ansatz should be regarded not only as potential
sources of destabilisation, but also as possible ingredients for new
controlled inflationary dynamics. It would also be interesting to extend this analysis to other LVS inflationary models, such as blow-up and poly-instanton inflation, to determine whether $\alpha^\prime$-loop corrections can significantly modify their inflationary potentials. In particular, logarithmically enhanced terms may play a more prominent role in these setups and could potentially generate ultra-slow-roll phases. A detailed study of these possibilities is left for future work.

\section*{Acknowledgments}
We would like to thank Xingyou He, Hong Lv, Simon Schreyer and Pramod Shukla for helpful discussions. XG and ZL are supported by the National Natural Science Foundation of China under Grant No.~12375065 and the Natural Science Foundation of Sichuan Province under Grant No.~2026NSFSC0033. C.W. is supported by the program PIPF-2024/TEC-34293 from Comunidad de Madrid.

\appendix

\section{String and Einstein frames} \label{frame}

\subsection*{From string frame to ten-dimensional Einstein frame}

The bosonic part of the type IIB supergravity action in the string frame is given by \cite{Blumenhagen:2013fgp,Polchinski:1998rr}\footnote{See p.~625 of \cite{Blumenhagen:2013fgp} and p.~90 of \cite{Polchinski:1998rr} for details.}
\begin{align}
	S^S_{\, IIB} =& \frac{1}{2  \tilde{\kappa}_{10}^2} \int d^{10}x \sqrt{-G^S} \Bigg\{ e^{-2\Phi} \left[ \mathcal{R}^S + 4(\nabla \Phi)^2 - \frac{|H_3|^2}{2} \right] - \left( \frac{|F_1|^2}{2} + \frac{|F_3|^2}{2} + \frac{|F_5|^2}{4} \right) \Bigg\}  \nonumber \\
	&- \frac{1}{4 \tilde{\kappa}_{10}^2} \int C_4 \wedge H_3 \wedge F_3  , \label{eq:IIB-string}
\end{align}
where \(l_s=2\pi\sqrt{\alpha'}\) and we have
\begin{equation}
     2\tilde{\kappa}_{10}^2=(2\pi)^7\alpha'^4
    =
    \frac{l_s^8}{2\pi},
\end{equation}

The standard Einstein-frame metric is defined by the Weyl transformation
\begin{align}
	g^S_{\,MN} = e^{\Phi/2} g^E_{\, MN} \,, \qquad
	g^{S\,MN} = e^{-\Phi/2} g^{E\, MN} \label{eq:string-to-einstein} \,.
\end{align}
In $D$-dimensional spacetime, the determinant of the metric then transforms as
	$\det[g^S] = e^{D\Phi/2}\det[g^E] $ and gives
	$\sqrt{-g^S}=e^{5\Phi/2}\sqrt{-g^E}$ in $D=10$.
Under a general conformal transformation of the form $\tilde{G}_{MN}=e^{2\omega}G_{MN}$, the Ricci scalar transforms as \cite{Tong:2009np}
\begin{equation}
\label{eq:ricci}
	\tilde{\mathcal{R}} = e^{-2\omega} \left\{ \mathcal{R} - 2 (D - 1) \nabla ^2 \omega - (D - 2)(D - 1) (\nabla \omega)^2 \right\},
\end{equation}
which in $D=10$  with $2\omega=\Phi/2$ leads to
\begin{equation}
	\mathcal{R}^S
	=
	e^{-\Phi/2}
	\left[
	\mathcal{R}^E
	-\frac{9}{2}\nabla^2\Phi
	-\frac{9}{2}(\nabla\Phi)^2
	\right],
\end{equation}
The total derivative term $-\frac{9}{2}\nabla^2\Phi$ can be dropped  since it does not contribute to the equations of motion after integration by parts. 
Therefore, the action \eqref{eq:IIB-string} in the Einstein frame becomes
\begin{align}
	S^E_{\, IIB}
	=& \frac{1}{(2 \pi)^7 \alpha^{\prime 4}}
	\int d^{10}x \sqrt{-G^E} e^{5\Phi/2}
	\nonumber\\
	&\quad \times \Bigg\{
	e^{-2\Phi}
	\Big[
	e^{-\Phi/2}\mathcal{R}^E
	- e^{-\Phi/2}\frac{9}{2}(\nabla \Phi)^2
	+ 4 e^{-\Phi/2}(\nabla \Phi)^2
	- \frac{e^{-3\Phi/2}|H_3|^2}{2}
	\Big]
	\nonumber\\
	&\hspace{1.5cm}
	- e^{-\Phi/2}\frac{|F_1|^2}{2}
	- e^{-3\Phi/2}\frac{|F_3|^2}{2}
	- e^{-5\Phi/2}\frac{|F_5|^2}{4}
	\Bigg\}
	\nonumber\\
	&\quad
	- \frac{1}{4 \tilde{\kappa}_{10}^2}
	\int C_4 \wedge H_3 \wedge F_3 .
\end{align}
with $|C_p^S|^2= e^{- p \Phi/2} |C_p^E|^2 $ for general $p$-form.
Combining the dilaton-dependent factors, one arrives at the standard type IIB supergravity action in the Einstein frame:
\begin{align}
	S^E_{\, IIB}
	=& \frac{1}{(2 \pi)^7 \alpha^{\prime 4}}
	\int d^{10}x \sqrt{-G^E}
	 \left\{
	\mathcal{R}^E
	- \frac{\partial_M\tau \partial^M \bar{\tau}}{2(\text{Im}\,\tau)^2}
	- \frac{G_3 \cdot \bar{G}_3}{2\text{Im}\,\tau}
	- \frac{|F_5|^2}{4}
	\right\}+ S_{CS},
	\label{IIB_Einstein}
\end{align}
where
\begin{equation}
	\tau = C_0 + \text{i}e^{-\Phi},
	\qquad
	G_3 = F_3 - \tau H_3, \qquad S_{CS}=\frac{1}{8 \text{i} \tilde{\kappa}_{10}^2}
	\int \frac{1}{\text{Im}\,\tau} C_4 \wedge G_3 \wedge \bar{G}_3 
\end{equation}
The Chern-Simons term, which we will neglect in the following, is topological and is not affected by the metric Weyl rescaling.  ~\eqref{IIB_Einstein} is the ten-dimensional frame used in the main text.

\subsection*{Four-dimensional Einstein frame and Planck units}

We now summarise the frame convention used in the main text. For the
purpose of fixing normalisations, we work in the unwarped orientifold
limit, take the axio-dilaton to be constant and set \(F_5=0\). The
relevant part of the ten-dimensional Einstein-frame action is
\begin{equation}
    S^E_{\rm IIB}
    \supset
    \frac{1}{2 \tilde{\kappa}_{10}^2}
    \int d^{10}x\sqrt{-g_{10}^E}
    \left[
        \mathcal R_{10}^E
        -
        \frac{G_3\cdot\bar G_3}{2\,{\rm Im}\,\tau}
    \right].
    \label{eq:10d-action-frame}
\end{equation}
Set ${\rm Vol}_6^E=\mathcal V_s\,l_s^6$
be the internal Einstein-frame volume with \(\mathcal V_s\) dimensionless.
Before the final four-dimensional Weyl rescaling, dimensional reduction
gives
\begin{equation}
    S_4
    \supset
    \frac{2\pi}{l_s^2}
    \int d^4x\sqrt{-g_4^E}\,
    \mathcal V_s\,\mathcal R_4^E
    -
    \frac{2\pi}{l_s^8}
    \int d^4x\sqrt{-g_4^E}\,
    \int d^6y\sqrt{g_6^E}\,
    \frac{G_3\cdot\bar G_3}{2\,{\rm Im}\,\tau} .
    \label{eq:4d-before-weyl}
\end{equation}

The four-dimensional Einstein-frame metric \(\tilde g_{\mu\nu}^E\) is
defined by the Weyl rescaling
\begin{equation}
    g_{\mu\nu}^E
    =
    \frac{\mathcal V_s^0}{\mathcal V_s}\,
    \tilde g_{\mu\nu}^E ,
    \qquad
    \mathcal V_s^0\equiv \langle \mathcal V_s\rangle ,
    \label{eq:weyl-rescaling-frame}
\end{equation}
where \(\mathcal V_s^0\) is the reference volume at the stabilised
vacuum. The Weyl rescaling also generates kinetic terms for the volume
modulus; these are part of the K\"ahler-moduli metric and are not
displayed in this normalisation discussion. From (\ref{eq:ricci}) we get 
\begin{equation}
     \sqrt{-g_4^E}\,\mathcal R_4^E
    =
    \frac{\mathcal V_s^0}{\mathcal V_s}
    \sqrt{-\tilde g_4^E}\,\tilde{\mathcal R}_4^E
    +\cdots ,
\end{equation}  
and the 4D Einstein--Hilbert term should match the first term of (\ref{eq:4d-before-weyl}):    
\begin{align}
  &  S_{\rm EH}
    =
    \frac{M_p^2}{2}
    \int d^4x\sqrt{-\tilde g_4^E}\,
    \tilde{\mathcal R}_4^E 
   = \frac{M_p^2}{2} \frac{\cV_s}{\cV_s^0}
    \int d^4x\sqrt{- g_4^E}\,
    {\mathcal R}_4^E 
    = \frac{2\pi}{l_s^2}
    \int d^4x\sqrt{-g_4^E}\,
    \mathcal V_s\,\mathcal R_4^E,
    \label{eq:Mp-frame}
\end{align}
which leads to
\begin{equation}
  M_p^2
    =
    \frac{4\pi\mathcal V_s^0}{l_s^2} = \frac{1}{8 \pi G }.
    \label{eq:M_p}
\end{equation}
This is the relation between the reduced 4D Planck mass and the stabilised
Calabi--Yau volume in the convention used in the main text.

After the same Weyl rescaling, the flux term  of (\ref{eq:4d-before-weyl}) takes the form
\begin{equation}
    S_{\rm flux}
    =
    -\frac{M_p^2}{2}
    \int d^4x\sqrt{-\tilde g_4^E}\,
    \frac{\mathcal V_s^0}{\mathcal V_s^2\,l_s^6}
    \int d^6y\sqrt{g_6^E}\,
    \frac{G_3\cdot\bar G_3}{2\,{\rm Im}\,\tau}.,
    \label{eq:Sflux-U}
\end{equation}
Equivalently, the four-dimensional scalar potential then follows as:
\begin{equation}
    V_{\rm flux}
    =
    \frac{M_p^2}{2}
    \frac{\mathcal V_s^0}{\mathcal V_s^2\,l_s^6}
    \int d^6y\sqrt{g_6^E}\,
    \frac{G_3\cdot\bar G_3}{2\,{\rm Im}\,\tau}.
    = 
    -\,i\,
    \frac{M_p^2}{2}
    \frac{\mathcal V_s^0}{\mathcal V_s^2\,l_s^6}
    \int
    \frac{G_3^+\wedge\bar G_3^+}{2\,{\rm Im}\,\tau},
    \label{eq:Vflux-frame}
\end{equation}
where the second equality follows from the standard decomposition into imaginary self-dual and anti-self-dual components; with our conventions, the combination $-\text{i}\int G_3^+\wedge\bar G_3^+$ is positive definite.\footnote{In the present analysis, we neglect warping effects and therefore set the warp factor $H=1$; see Section~4 of \cite{ValeixoBento:2023afn} for a more detailed discussion of the warped case.}
In the main text we set \(M_p=1\) after this rescaling.

\subsection*{Relation to the \texorpdfstring{\(\mathcal N=1\)}{N=1} F-term potential}

We now relate the flux energy obtained from dimensional reduction to the
standard four-dimensional \(\mathcal N=1\) supergravity expression. We
define the dimensionless Gukov--Vafa--Witten superpotential by
\begin{equation}
    W_{\rm GVW}
    =
    \frac{1}{l_s^2}
    \int_{X_6} G_3\wedge \Omega .
    \label{eq:GVW-dimensionless}
\end{equation}
With this normalization, the positive flux contribution can be written in
terms of the F-terms of the axio-dilaton and complex-structure moduli as
\begin{equation}
    -\,i
    \int_{X_6}
    {G_3^+\wedge \bar G_3^+}
    =
    \frac{2\, l_s^4}
         {
          i\int_{X_6}\Omega\wedge\bar\Omega}
    K^{\alpha\bar\beta}
    D_\alpha W_{\rm GVW}
    D_{\bar\beta}\bar W_{\rm GVW}.
    \label{eq:Gplus-Fterms}
\end{equation}
Here \(\alpha,\bar\beta\) run over the axio-dilaton and complex-structure
moduli only. The precise numerical normalization depends on the convention
for the form norm and  \(\Omega\); these convention-dependent factors
will be absorbed into the normalization of the four-dimensional
superpotential below.

The tree-level K\"ahler potential before integrating out the
axio-dilaton and complex-structure moduli is
\begin{equation}
    \frac{K}{M_p^2}
    = K_0 + K_{\tau}+K_{cs} =
    -2\ln\mathcal V_s
    -\ln\!\left(2\,{\rm Im}\,\tau\right)
    -\ln\!\left(
        \frac{i}{l_s^6}
        \int_{X_6}\Omega\wedge\bar\Omega
    \right).
    \label{eq:Ktree-full}
\end{equation}
The factor \(l_s^{-6}\) only makes the argument of the logarithm
dimensionless and can equivalently be absorbed into the normalization of
\(\Omega\).

Using the no-scale identity for the K\"ahler moduli
$K^{i\bar j}K_iK_{\bar j}=3$,
the positive flux potential can be rewritten in the full moduli notation as
\begin{align}
	V_{\rm flux}
	=&
	\frac{M_p^2}{2}
	\frac{\mathcal{V}^0_s}{\mathcal{V}^2_s l^6_s}
	\frac{1}{2 \,\text{Im}\,\tau}\frac{2 \, l_s^4}{ i \,\int \Omega \wedge \bar \Omega}
	\left(
	K^{I \bar J} D_I W_{\rm GVW} D_{\bar J} \bar W_{\rm GVW}
	- 3 |W_{\rm GVW}|^2
	\right)
	\nonumber\\
	=&
	\frac{M_p^4 \mathcal{V}^0_s}{l^{2}_s}
	e^{-2 \ln{\mathcal{V}_s}}
	e^{-\ln\left(2\text{Im}\,\tau\right)}
	e^{-\ln\left(\text{i}\int \Omega \wedge \bar \Omega\right)}
	\left(
	\frac{K^{I \bar J}}{M_p^2}D_I W_{\rm GVW} D_{\bar J}\bar W_{\rm GVW}
	- \frac{3}{M_p^2}|W_{\rm GVW}|^2
	\right)
	\nonumber\\
	=&
	\frac{M_p^6}{4 \pi}
	e^{K/M_p^2}
	\left[
	K^{I \bar J} D_I W_{\rm GVW} D_{\bar J} \bar W_{\rm GVW}
	- \frac{3}{M_p^2}|W_{\rm GVW}|^2
	\right],
	\label{eq:v-flux-e}
\end{align}
where \(I,\bar J\) now run over all chiral multiplets. This is the
standard F-term structure, up to the overall prefactor inherited from the
ten-dimensional normalization and the dimensionless definition of \(W_{\rm GVW}\).

After flux stabilization, the axio-dilaton is fixed at
\begin{equation}
    \langle{\rm Im}\,\tau\rangle=e^{-\Phi_0}=\frac{1}{g_s},
    \qquad
    e^{K_\tau}
    =
    \frac{1}{2\,{\rm Im}\,\tau}
    =
    \frac{g_s}{2}.
    \label{eq:dilaton-fixed-appA}
\end{equation}
The complex-structure moduli are also fixed by the flux.
Thus \(e^{K_{\rm cs}}\) is a constant in the K\"ahler-moduli effective
theory and can be absorbed into the definition of the flux
superpotential,
\begin{equation}
    W
    \equiv
    e^{K_{\rm cs}/2}W_{\rm GVW},
    \qquad
    W_0
    \equiv
    \langle W\rangle .
    \label{eq:W-rescale-Kcs}
\end{equation}
With this convention, the K\"ahler-sector potential becomes
\begin{equation}
    V_F
    =
    \frac{g_sM_p^6}{8\pi}\,
    e^{K_0}
    \left[
        K^{i\bar j}D_iW D_{\bar j}\overline W
        -
        \frac{3}{M_p^2}|W|^2
    \right],
    \label{eq:VF-before-Mp-one}
\end{equation}
where \(i,\bar j\) now run over the K\"ahler moduli. Finally, in the
main text we set \(M_p=1\), replace \(K_0\) by the corrected K\"ahler
potential \(K\), and write
\begin{equation}
    V_F
    =
    \kappa e^K
    \left(
        K^{i\bar j}D_iW D_{\bar j}\overline W
        -
        3|W|^2
    \right),
    \qquad
    \kappa\equiv\frac{g_s}{8\pi}.
    \label{eq:main-kappa-convention-appA}
\end{equation}
This is the convention used throughout the main text. In particular,
the symbol \(W\) there denotes the dimensionless four-dimensional
superpotential after the constant complex-structure factor has been
absorbed into \(W_0\)\footnote{
One may alternatively define a dimensionful superpotential
\(
\widehat W=M_p^3W_{\rm GVW}/\sqrt{4\pi}
\).
In that convention the factor \(M_p^6/(4\pi)\) in
\eqref{eq:v-flux-e} is absorbed into \(\widehat W\), and the F-term
potential takes the standard supergravity form without the explicit
prefactor \(\kappa\). We do not use that convention in the main text.
}.

\subsection*{Modified Einstein frame}

The main text uses the standard Einstein frame. Some
LVS literature, however, uses a modified Einstein frame, and we briefly
summarize the relation in order to avoid confusion about powers of
\(g_s\). The modified Einstein-frame metric is defined by
\begin{equation}
    g^S_{MN}
    =
    e^{(\Phi-\Phi_0)/2}g^{\rm ME}_{MN},
    \qquad
    \Phi_0=\langle\Phi\rangle ,
    \label{eq:modified-frame-def}
\end{equation}
so that at the physical vacuum, \(\Phi=\Phi_0\), the string-frame metric
coincides with the modified Einstein-frame metric.  After dilaton stabilization, comparing with the
standard Einstein-frame relation
$
    g^S_{MN}=e^{\Phi/2}g^E_{MN},
$
one obtains,
\begin{equation}
    g^{\rm ME}_{MN}
    =
    e^{\Phi_0/2}g^E_{MN}
    =
    g_s^{1/2}g^E_{MN}.
    \label{eq:ME-standard-relation}
\end{equation}
Thus the two frames differ only by a constant Weyl rescaling once the
dilaton is fixed.

From the string-frame action, the ten-dimensional action in the
modified frame contains an overall factor \(e^{-2\Phi_0}=g_s^{-2}\).
Consequently, dimensional reduction gives
\begin{equation}
    M_{p,{\rm ME}}^2
    =
    \frac{4\pi\,\mathcal V_s^0}{g_s^2\,l_s^2},
    \label{eq:Mp-ME}
\end{equation}
instead of the standard Einstein-frame relation \eqref{eq:M_p}.
Using the same dimensionless GVW superpotential
\eqref{eq:GVW-dimensionless} and the rescaling in \eqref{eq:W-rescale-Kcs},
the F-term potential becomes
\begin{equation}
    V_F^{\rm ME}
    =
    g_s^3\,\kappa\,
    e^K
    \left(
        K^{i\bar j}D_iW D_{\bar j}\bar W
        -
        3|W|^2
    \right),
    \qquad
    \kappa\equiv\frac{g_s}{8\pi},
    \label{eq:VF-ME}
\end{equation}
where we have set \(M_p=1\) in the last expression.
The extra factor \(g_s^3\) in \eqref{eq:VF-ME} is not a physical effect;
it simply reflects the modified-frame normalisation.

\section{\texorpdfstring{Origin of the $\alpha^\prime$-loop corrections}{Origin of the alpha-prime-loop corrections}}
\label{app:alpha_loop_origin}

This appendix follows \cite{Gao:2022uop} to justify the
minimal ansatz \eqref{eq:deltaK-alpha-loop-fibred} and to clarify the
relation between the parameters $C_2,C_3,C_4$ and the perturbative effects
identified in the 10D EFT analysis.  

\subsection*{Genuine non-local loop effects}
Genuine loop effects arise from finite one-loop contributions of ten-dimensional or brane-localized fields propagating through the compact space. 
In a K3-fibred compactification it is useful to introduce two length
scales, $L_f$ for the fibre and $L_b$ for the base, with
\begin{equation}
    \mathcal V\sim L_b^2L_f^4,
    \qquad
    \tau_f\sim L_f^4,
    \qquad
    \tau_b\sim L_b^2L_f^2 .
    \label{eq:length-tau-relations}
\end{equation}
The two-step reduction can be organized as
$10d\xrightarrow{\tau_f}6d\xrightarrow{t_b}4d$.  
We first compactify the ten-dimensional theory on the fibre four-cycle
and integrate out the associated fibre KK tower. After the subsequent
reduction to four dimensions and Weyl transformation to the
four-dimensional Einstein frame, dimensional analysis gives kinetic
structures
\begin{equation}
    \Delta\mathcal L^{(f)}_{\rm kin}
    \sim
    \frac{(\partial\tau_b)^2}{\tau_b^2\tau_f^2}
    +
    \frac{(\partial\tau_b)(\partial\tau_f)}{\tau_b\tau_f^3}
    +
    \frac{(\partial\tau_f)^2}{\tau_f^4}
    +\cdots .
    \label{eq:fibre-loop-kinetic-structures}
\end{equation}
The last term is reproduced by correction
$\delta K_{gl}^{(f)}\supset 1 /\tau_f^2$, which is included in $C_2$.  The mixed kinetic
structures may formally be integrated to functions containing logarithms,
such as $\delta K_{gl}^{(f)}\supset \ln\tau_b/\tau_f^2$.  However, such functions also generate 
additional terms in the kinetic metric, and without a complete
Calabi--Yau calculation it is not possible to decide whether the unwanted
pieces cancel or are removed by field redefinitions.  The minimal ansatz
therefore retains the unambiguous fibre-scale structure $C_2/\tau_f^2$ where $C_2$ denotes the fibre-scale degree $(-2)$ coefficient.

Fields localised on D7-branes can also generate genuine loop
corrections. Their moduli dependence depends on the wrapped divisor and
on the KK spectrum of the worldvolume theory. For a D7-brane wrapping
the fibre, the resulting
correction has the same $ \delta K_{gl}^{D7(f)}\supset 1/(\cV \sqrt{\tau_f}) \sim 1/(\tau_b\tau_f)$ dependence and  hence contributes to the mixed coefficient $C_3$. 
By
contrast, loops on a D7-brane wrapping the base-associated four-cycle give $ \delta K_{gl}^{D7(b)}\supset 1/\tau_f^2$ and
need not introduce a parametrically new structure and can renormalise
the fibre-sector contribution represented by \(C_2\).
Thus \(C_2\) and \(C_3\) label moduli structures rather than simply the cycle wrapped by the D7-brane. 

Starting from the six-dimensional EFT obtained after the first
compactification step, we next compactify on the base and integrate out
the corresponding KK tower. The resulting four-dimensional kinetic
structures have the schematic form
\begin{equation}
    \Delta\mathcal L^{(b)}_{\rm kin}
    \sim
    \frac{(\partial\tau_b)^2}{\tau_b^4}
    +
    \frac{(\partial\tau_b)(\partial\tau_f)}
         {\tau_b^3\tau_f}
    +
    \frac{(\partial\tau_f)^2}
         {\tau_b^2\tau_f^2}
    +\cdots .
    \label{eq:base-loop}
\end{equation}
The first term is reproduced by $\delta K_{\rm gl}^{(b)} \supset  {1}/{\tau_b^2}\sim  {\tau_f}/{\mathcal V^2}$ and is represented by the coefficient \(C_1\).  In BHP-driven configurations, (\ref{eq:base-loop}) can therefore
be absorbed into the effective coefficient of the corresponding
growing term. In the minimal BHP-free analysis we neglect these terms  in the large volume limit.

\subsection*{Local D7/O7 corrections}
Local $\alpha'$ corrections are associated with higher-derivative
operators in the ten-dimensional bulk, on D-branes/O-planes, or on their
intersection loci.  
First, relevant localized operators such as $R_8^2$ term on
D7/O7 systems, generate K\"ahler potential corrections
homogeneous of degree $(-1)$.  These corrections are not part of
\eqref{eq:deltaK-alpha-loop-fibred} since  they are subject to the extended
no-scale cancellation and contribute to the scalar potential at the same
parametric order as KK-type BHP corrections, but through the quadratic term
in the expansion in $\delta K$.  In the main text they are included in the BHP or local-KK baseline rather than in $V_{\rm \alpha'-loop}$.
Second, marginal localized operators may generate logarithmic running.  For
$ R_8^4$
   on a D7/O7 worldvolume, its coefficient can run between the string scale
and the compactification scale, producing a factor
$
    c_0+c_{\log}\ln(M_{10}g_s^{1/4}L) .
$
For a D7/O7 system wrapping the base-associated four-cycle of a K3-fibred
geometry, the first step of the reduction can give
\begin{equation}
    \ln(M_{10}g_s^{1/4}L_f)
    =
    \frac14\ln\tau_f+{\rm constant},
    \label{eq:log-Lf-tauf}
\end{equation}
where the constant is absorbed into a finite threshold coefficient.  The
dimensional reduction of the unknown tensor contractions of $R_8^4$ can
produce several moduli structures.  The minimal mixed term relevant for
fibre inflation is
\begin{equation}
    \delta K_{\rm loc}^{\rm mixed}
    \supset
    \frac{C_3+C_4\ln \tau_f}{\tau_b\tau_f},
    \label{eq:local-mixed-c3c4}
\end{equation}
where $C_3$ includes the finite threshold accompanying the same localized
sector, while $C_4$ is the coefficient of the logarithmic running.  Similar
logarithmic contributions may arise from candidate marginal $R_6^3$
operators localized on D7/O7 intersection loci.
This avoids identifying $C_3$ solely with a relevant D-brane operator.
Relevant operators give degree $(-1)$ corrections; the mixed degree $(-2)$
coefficient $C_3$ is instead the finite part of the retained D7/O7-sector
threshold or higher-derivative correction.

As a summary, the effective coefficients retained in
\eqref{eq:deltaK-alpha-loop-fibred} therefore have the schematic
interpretation. $C_1$ represents base-sector degree $(-2)$ genuine-loop structure, $C_2$ represents fibre-sector bulk and brane-localised genuine loops, $C_3$ represents finite mixed genuine-loop and local-threshold pieces and $C_4$ represents logarithmic running of candidate marginal local operators.

\bibliographystyle{utphys}
\bibliography{ML}

\end{document}